# Classification of Metal - Insulator Transitions:

# Relating characteristic Properties to Quantum Chemical Bonding Descriptors

**Tim Bartsch[1], Carl-Friedrich Schön[1], Dasol Kim[1],**

**Raagya Arora[2], Umesh Waghmare[3], Matthias Wuttig[1,4,5*]**

[1] I. Institute of Physics, Physics of Novel Materials, RWTH Aachen University, 52056 Aachen, Germany

[2] John A. Paulson School of Engineering and Applied Sciences, Harvard University, Cambridge, MA 02138, USA

[3] Theoretical Sciences Unit, School of Advanced Materials JNCASR Jakkur, Bangalore 560064, India

[4] Jülich-Aachen Research Alliance (JARA FIT and JARAHPC), RWTH Aachen University, 52056 Aachen, Germany

[5] Green IT (PGI 10), Forschungszentrum Jülich GmbH, 52428 Jülich, Germany

*Corresponding author. Email: wuttig@physik.rwth-aachen.de

## Abstract:

Pressure-induced metal–insulator transitions (MITs) are classified by the evolution of characteristic opto-electronic and vibrational properties calculated with density functional theory. Three classes emerge: ionic solids metallize continuously at band-gap closure with hardening phonons; covalent solids show discontinuous changes in atomic arrangement and optical phonon frequencies; a third class exhibits complete lattice softening and drastically enhanced electron–phonon coupling. A one-dimensional hydrogen chain reproduces this behavior and serves as a toy model of the underlying bonding mechanism, termed metavalent. Two quantum-chemical descriptors capture the distinct bonding changes behind the three classes. In metavalent solids, competing electron localization and delocalization yield soft optical modes and Peierls distortions on the insulating side, superconductivity on the metallic side, and low lattice thermal conductivity near the MIT.

## Introduction: The Origin of Metal – Insulator Transitions

The origin and nature of the transition between metals and insulators is one of the most fascinating topics in condensed matter physics and remains an active field—for example, in efforts to metalize hydrogen under extreme pressures[1,2]. The end points of the transition are well defined: a solid with non-vanishing electrical conductivity σ for T → 0 K is a metal, while an insulator has σ = 0 (T → 0 K).

Six decades ago, the nature of the MIT was debated, particularly with regard to **how** the insulating state is reached. Mott claimed a minimum metallic conductivity below which no metallic state can exist[3,4], whereas Abrahams et al. proposed a second-order quantum phase transition at σ = 0 and T = 0 K[5,6]; Rosenbaum et al. later provided experimental evidence for a continuous conductivity change in phosphorus-doped Si [7,8], supporting the modern view. The **origin** of this MIT, however, remains controversial: competing arguments favor correlation or disorder[9], and since the two are hard to disentangle, the MIT in doped semiconductors is now considered a Mott–Anderson MIT, governed by both[7,8,10]. This ambiguity raises a fundamental question: can MITs be classified by their characteristic accompanying property changes, and would such changes provide insight into their origin? The present work addresses a different class of transitions. Unlike Mott transitions, driven by strong electron–electron correlation, or Anderson transitions, driven by disorder-induced localization, the MITs investigated here occur in ordered crystalline solids and are driven by pressure-induced changes in band structure and chemical bonding. The key questions are whether these pressure-controlled MITs can be classified by unique changes in their characteristic properties, and how such a classification can be explained.

With this in mind, we have examined the properties of various solids undergoing pressure-induced MITs. As the insulators become metallic upon increasing pressure, the key questions answered here are whether these pressure-controlled MITs can be classified based on unique changes in their characteristic properties and how such a classification can be explained.

**Classifying characteristic property changes in different MITs**

To this end, several properties have been studied which are expected to vary systematically with pressure. These include properties that quantify the electronic subsystem, such as the band gap $E_G$ and the optical dielectric constant $\varepsilon_\infty$, the electronic polarization induced by a vibrational displacement is characterized by the Born effective charge $Z_+^*$, and properties of the lattice (in particular, the frequencies of the longitudinal and transverse optical modes ($\omega_{LO}$ and $\omega_{TO}$), as well as the Grüneisen parameter $\gamma_{LO}$ of the optical mode, a measure of anharmonicity. Vanishing of the band gap marks the point of metal-insulator transition (MIT) and is indicated by a dashed blue line in the corresponding figures. These quantities can also be viewed as complementary response functions that probe different aspects of the material. The optical dielectric constant describes the electronic response, the Born effective charge characterizes the coupled electronic polarization-lattice response, while the optical phonon frequencies and the corresponding Grüneisen parameter quantify the lattice response to external perturbations. Their systematic evolution under compression provides a characteristic fingerprint of the different classes of metal-insulator transitions and forms the basis for the classification presented in this work. To facilitate comparison, we plot the transition as a function of the resulting volume rather than pressure. The relationship between pressure and volume of each solid discussed is depicted in the SI (Figures S7-10).

Figure 1 A-E shows changes of properties upon the MIT for NaCl, which serves as a prototype of ionic bonding. NaCl undergoes a transition from the $Fm\bar{3}m$ phase to the $Pm\bar{3}m$ structure in our simulations, when the pressure is increased to 27.9 GPa, consistent with literature, where values of around 30 GPa[11,12], and becomes metallic at 573 GPa. The MIT is continuous: the band gap narrows steadily until it closes (Figure 1 A), accompanied by a concomitant increase in electronic polarizability, as measured by $\varepsilon_\infty$ (Figure 1 B). In contrast, the Born effective charge $Z^*_+$ (normalized by the formal oxidation state), which probes the coupling between atomic displacements and electronic polarization and thus measures the chemical bond polarizability, remains almost constant around 1 (Figure 1 C), while the LO phonon frequency (Figure 1 D) hardens continuously without any anomaly at the MIT. NaCl may undergo its MIT in a related oC8 phase[13], yet with almost identical property changes (Figure S2), and other ionic compounds such as CsF and CsCl behave alike (Figure S1 & S2). Hence, in all ionic solids studied only the electronic subsystem

responds to electron delocalization, whereas the lattice ($\omega_{LO}$, $\gamma_{LO}$) and the change in polarization induced by an atomic displacement ($Z^*_+$) show no notable change at the MIT.

Covalent solids behave differently. III–V and group IV semiconductors undergo a first-order, volume-discontinuous transition to a high-pressure structure[14-16], which for GaAs, Si, Ge, InSb, InAs, InP, AlSb, AlP, AlAs and GaP is accompanied by an MIT[14,15,17-19]. For GaAs (Figure 1 K–P), Si and Ge (Figure S3) the first-order nature of the transition leads to a distinct property change at the MIT: all high-pressure structures are metallic (Table S1), so that $E_g$ jumps discontinuously to zero (Figure 1 K), unlike the continuous closure in ionic solids (a discussion of the DFT band-gap problem in the context of this study is provided in Figure S5 and Table S2.). $Z^*_+$ remains nearly constant as the MIT is approached (Figure 1 M), and the LO mode (Figure 1 N) shows no precursor of the transition either: it increases almost linearly in the zinc blende phase, drops discontinuously at the transition to the rock salt structure and rises again. Accordingly, the mode Grüneisen parameter (Figure 1 P) shows only a small, discontinuous change. Other covalent solids resemble GaAs (Figure S3). Ionic and covalent solids can thus be distinguished by how they reach the metallic state – raising the question of whether a further class of MITs exists, characterized by a distinctly different property change upon electron delocalization.

GeTe indeed represents such a class (Figure 1 F–J). As in ionic compounds, the band gap narrows continuously until it closes (Figure 1 F), and $\varepsilon_\infty$ diverges at the MIT (Figure 1 G) – yet far more strongly than in ionic solids. Even more striking is the electron–phonon coupling: $Z^*_+$ increases dramatically in the vicinity of the MIT (Figure 1 H), a behavior absent in ionic and covalent solids and evidence for a close link between electronic delocalization and electron–phonon coupling. The lattice is affected as well: the frequencies of both the TO and the LO mode go to zero at the MIT (Figure 1 I), i.e., the lattice softens completely, revealing a strong coupling between electronic and lattice degrees of freedom – clearly a distinctly different type of MIT. Consistently, the Grüneisen parameter (Figure 1 J), a measure of anharmonicity, increases dramatically near the MIT and exceeds 50, the signature of a highly anharmonic state. Qualitatively the same behavior is found for SnTe and GeSe (Figure S6). Hence, at least three classes of pressure-induced MITs exist; their characteristic property changes are compared in Table 1. We note in passing that the MIT in GeTe differs fundamentally from a ferroelectric transition as in $BaTiO_3$: GeTe is an improper ferroelectric[20], whose spontaneous polarization is driven by another primary order parameter, the Peierls distortion.

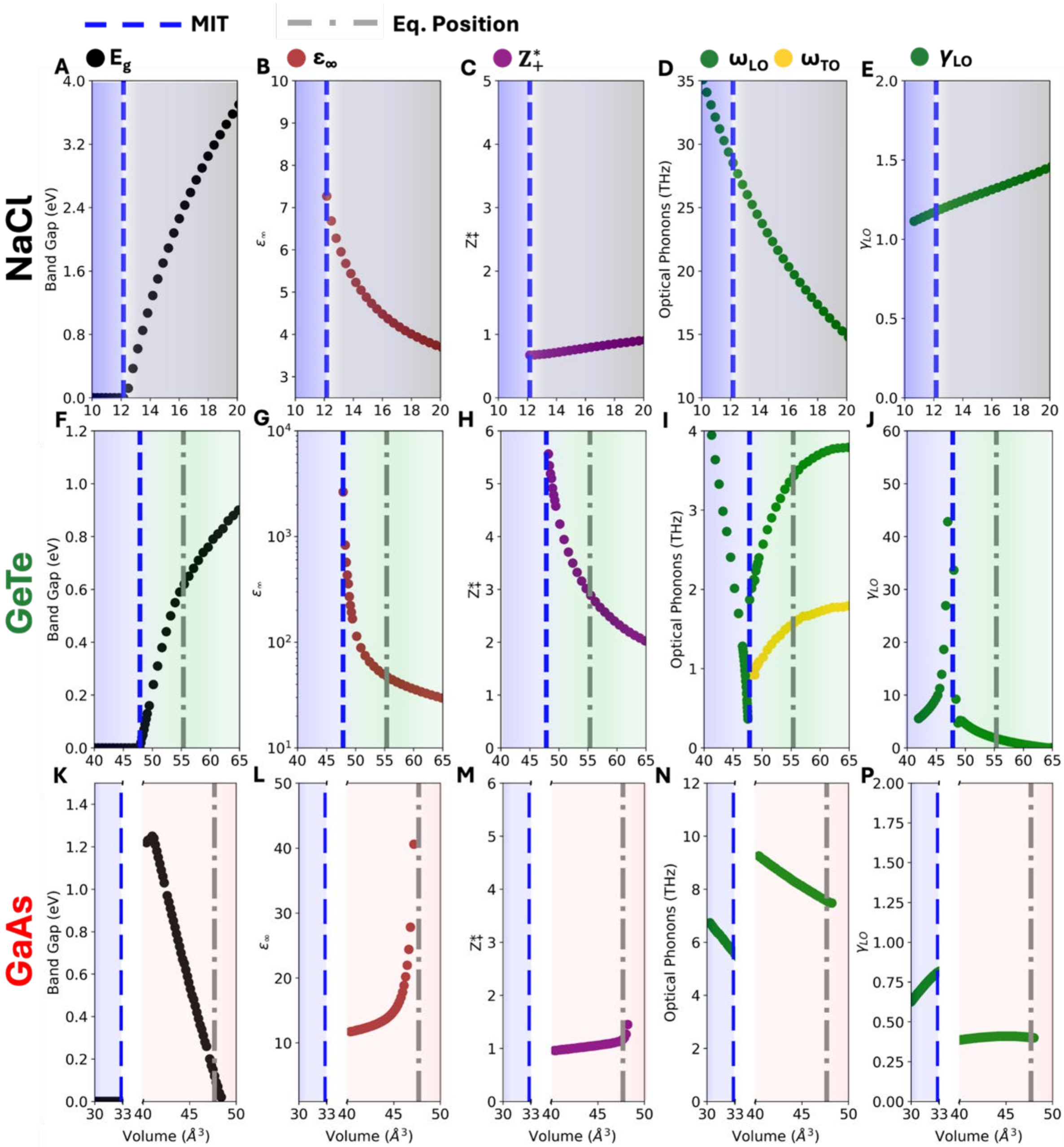

***Figure 1:*** *Evolution of material properties with cell volume for a representative compound of each bonding type. Ionic bonding is represented by NaCl (structure shown is the Pm3̅m phase) (A-E). GeTe is shown as a prototype for metavalent bonding (right of the blue vertical line: distorted cubic phase ($R\overline{3}m$); left of the blue line: Fm3̅m phase) (F-J). Covalent bonding is represented by GaAs (right of the blue line: zincblende phase; left of the blue line: Fm3̅m phase) (K-P). Upon continuous compression from the equilibrium structure (indicated by a dash-dotted grey line) to the metallic state (dashed blue line), a characteristic evolution of the corresponding property portfolio ($E_g$, $\varepsilon_\infty$, $Z^*_+$, $\omega_{LO}$ and $\omega_{TO}$) can be observed for each compound. The transition of the metavalent solid differs significantly from that of ionic and covalent solids, as it is characterized by a strong coupling of electronic and lattice properties. Please note that different scales have been chosen to maximize the clarity of the trends shown.*

*Table 1. Taxonomy of MITs based on characteristic changes in $E_g$, $\varepsilon_\infty$, $Z^*_+$, $\omega_{LO}$, $\gamma_{LO}$ as the MIT is approached from the insulating side, and at the transition itself, for ionic, metavalent and covalent solids. A detailed discussion of the categorization can be found in the SI.*

| | **Ionic MIT** | **Metavalent MIT** | **Covalent MIT** |
|---|---|---|---|
| **Solid** (Color identifies type of bonding in ground state) | NaCl (Pm3m), CsCl (Pm3m), CsF (Pm3m), NaCl (oC8), CsCl (Pbam), CsI*, CsBr*, BaS*, BaSe*, BaTe*, KI*, RbI*, GaN*, AlN*, InN*, PbTe | GeTe, SnTe, GeSe | GaAs, Si, Ge, InSb, InAs, InP, AlSb, AlP, AlAs, ZnSe*, ZnTe*, ZnS* |
| **Band Gap $E_g$** | continuous | continuous | discontinuous |
| **Dielectric Constant $\varepsilon_\infty$** | increases | diverges | discontinuous |
| **Born Effective Charge $Z^*_+$** **Coupled lattice-electronic polarization response** | unaffected by MIT | strong increase at MIT | discontinuous |
| **Longitudinal Optical Mode $\omega_{LO}$** | continuous hardening | lattice softens at MIT | discontinuous hardening |
| **Grüneisen parameter $\gamma_{LO}$, change at MIT** | small, continuous | diverges | small, discontinuous |

These findings offer a route to comprehend metavalent bonding via its MIT signature, provided a model system with the same characteristics can be identified. Interestingly, a one-dimensional hydrogen chain with two atoms per unit cell, in which the atoms can move freely along the chain axis, represents such a system; depending upon cell size, it exhibits two limiting cases, a covalent and a metallic phase.

In the molecular limit (large cell size), covalently bonded $H_2$ molecules form, with a vibrational frequency of 132 THz (4401 $cm^{-1}$)[21] (Figure 2 C) and a band gap of 10.14 eV[22] (Figure 2 A); at small cell sizes the chain is gapless. In between, a third region with remarkable properties[23] exists: the band gap narrows continuously (Figure 2 A), $\varepsilon_\infty$ increases concomitantly (Figure 2 B), the LO mode softens to a minimum at the MIT (Figure 2 C), where the mode

Grüneisen parameter diverges (Figure 2 D) – exactly the features of the metavalent MIT. Both MITs hence fall into the same property class, and the hydrogen chain can serve as a toy model of the metavalent bond.

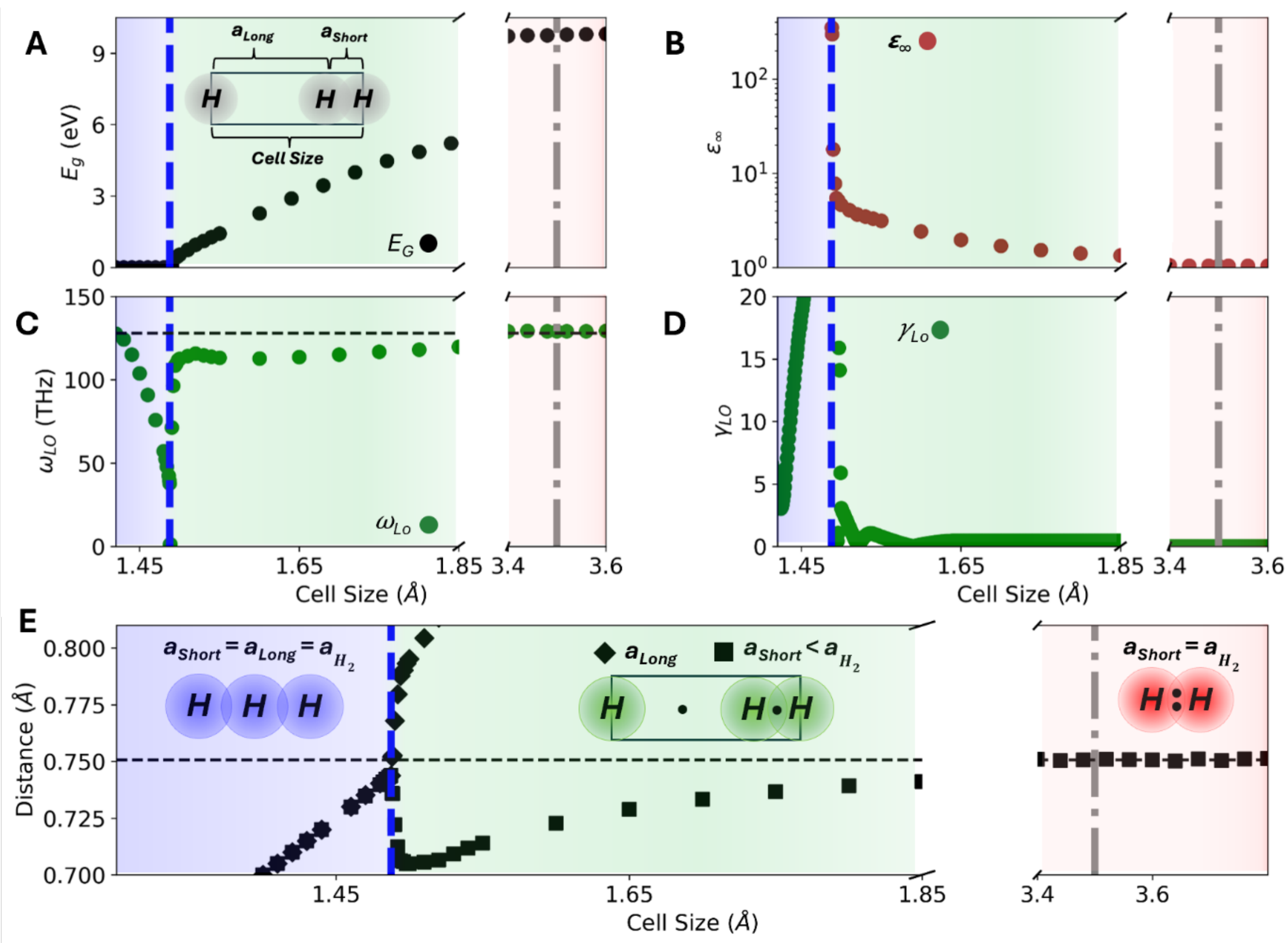

***Figure 2:*** *Evolution of characteristic properties of the one-dimensional hydrogen chain under compression. (A) progression of the bandgap for different cell sizes (B) optical dielectric constant $\varepsilon_\infty$, (C) longitudinal optical (LO) phonon mode and (D) Grüneisen parameter of the LO mode ($\gamma_{LO}$). The dash-dotted grey line indicates the most stable phase, while the dashed blue line marks the closure of the band gap. A horizontal black line represents the vibrational frequency of the isolated hydrogen molecule. Three characteristic bonding regimes are distinguished and color-coded: covalent (red), metavalent (green), and metallic (blue), as motivated in the main text.(E) Structural evolution of the 1-Dimensional hydrogen chain containing 2 atoms per unit cell depicted along the short and long bond. The dashed horizontal black line represents the characteristic $H_2$ bond distance of 0.741 Å.*

Structurally, the chain can likewise be divided into three regions (Figure 2 E). At large cell sizes it dimerizes into a molecular ground state with alternating $a_{short}$ = 0.75 Å and $a_{long}$ = 2.71 Å, found at a cell size of 3.46 Å (grey dash-dotted line)[24,25]. The opposite limit is the equidistant chain ($a_{short} = a_{long}$), which has no band gap independent of cell size and is stabilized at high pressure (left of the blue line in Figure 2 A). The intermediate region, where the significant property changes on the insulating side occur, is characterized by an unusual atomic arrangement: the short bond drops below the dimer distance ($a_{short} < a_{Dimer}$), even though the cell size exceeds twice the dimer distance.

So far, three classes of MITs – ionic, covalent and metavalent – have been distinguished solely based on the accompanying physical properties, and the metavalent class closely resembles the hydrogen chain near its MIT. These characteristics will now be related to the atomic arrangement and the bonds between adjacent atoms (Figure 3). Structural changes are quantified by the nearest-neighbor bond lengths and by the effective coordination number (ECoN)[26] , a distance-weighted average of the number of neighbors of a given site (Figure 3 B,F,J). To link structure to bonding, the numbers of electrons shared and transferred are evaluated (Figure 3 C,G,K); both have recently been shown to be quantum-chemical bonding descriptors with predictive power for properties such as the band gap, $Z^*_+$ or the Grüneisen parameter[23,27,28]. Finally, $\omega_{LO}$ is re-plotted (Figure 3 D,H,L) to connect structure, bonding and properties.

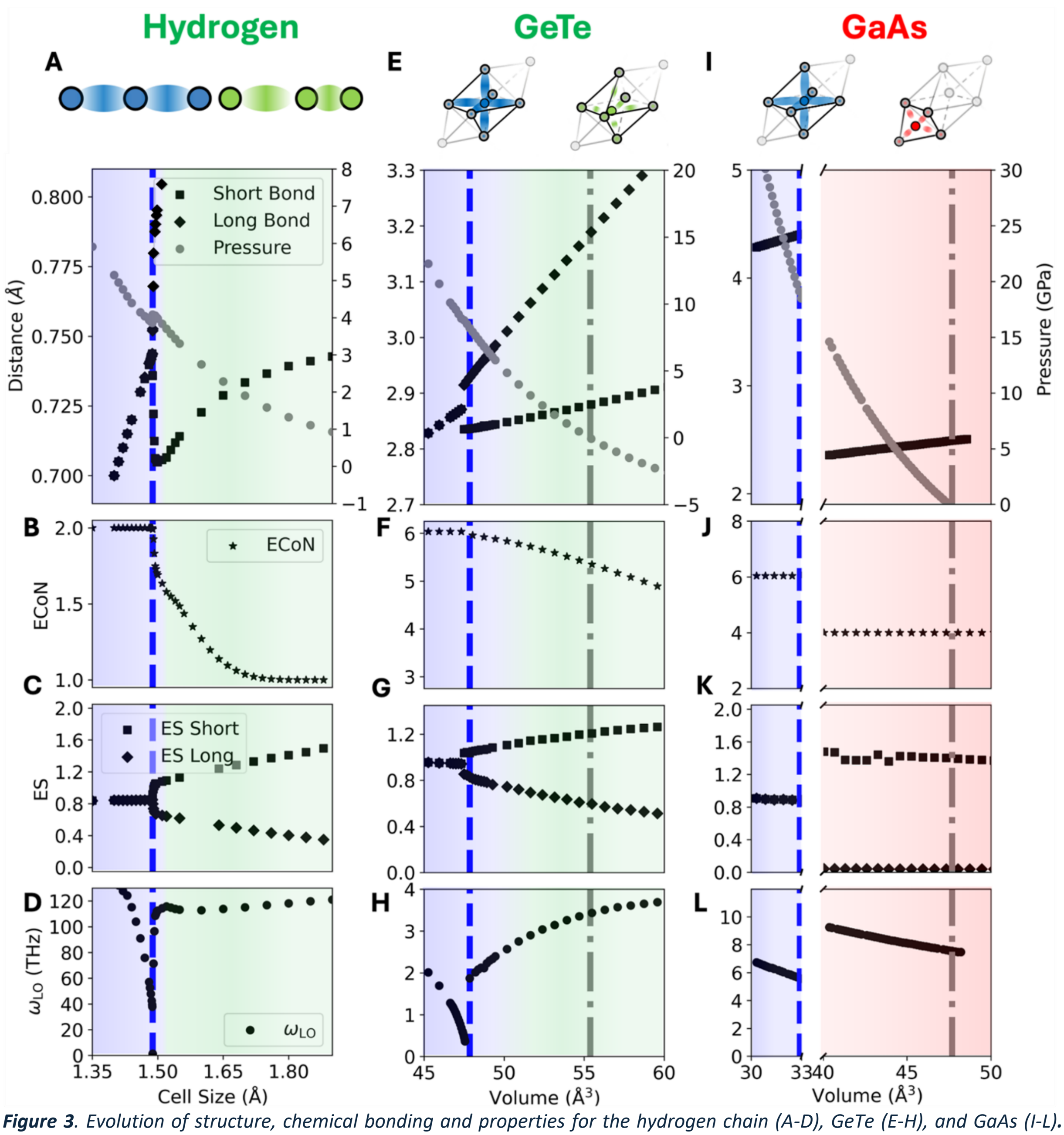

***Figure 3**. Evolution of structure, chemical bonding and properties for the hydrogen chain (A-D), GeTe (E-H), and GaAs (I-L)**.** To characterize the structural progression the short and long bond as well as the effective coordination number (ECoN) is depicted, together with the number of electrons shared between adjacent atoms and the frequency of the LO mode ($\omega_{LO}$) at the $\Gamma$-point. In the transition zone from metavalent to metallic bonding, the atoms in GeTe and the H chain adapt an unusually short distance. In this region, $\omega_{LO}$ is softest when the atomic arrangement just becomes symmetric and the ECoN approaches 6 (for GeTe) or 2 (for the H chain).*

Interestingly, the three MITs can also be discerned by their atomic arrangement at the transition. The ionic compounds show no discontinuous change of atomic positions upon compression: the ECoN stays constant at 11.47, ES increases continuously towards the MIT and ET remains constant apart from numerical noise. Since these structural changes are small, the corresponding data are shown in Figures S7–S8, and Figure 3 focuses on the difference between covalent and metavalent solids.

For GaAs (Figure 3 I) and other covalent solids (Figure S9), the MIT coincides with a first-order phase transition: the density increases suddenly and the ECoN jumps from 4 to 6 (Figure 3 J), in line with the discontinuous change of ES (Figure 3 K). Neither structure (Figure 3 I), bonding (Figure 3 K) nor properties (Figure 3 L) show any precursor of the MIT.

Metavalent solids (Figure 3 E–H, Figure S10), finally, show a peculiar behavior: right at the MIT, the short bond on the insulating side is even shorter than the bond on the metallic side, although the cell is larger (Figure 3 E). The $a_{long}/a_{short}$ ratio decreases with compression; ratios below 1.2 are associated with metavalent bonding (green background)[29,30]. The ECoN (Figure 3 F) takes non-integer values between full localization (ECoN = 3, in accordance with the 8 – N rule) and full delocalization (ECoN = 6)[28] , and the number of electrons shared approaches 1 close to the MIT (Figure 3 G). The hydrogen chain (Figure 3 A–D) shows the analogous region where the short bond drops below the dimer distance. There, the ECoN deviates from the molecular value of 1 and takes values below 2, violating the 2 – N rule for molecules, and the evolution of ES towards the MIT closely resembles that of GeTe. In both the hydrogen chain and GeTe, band-gap closure ($E_g \rightarrow 0$) is accompanied by a complete softening of the optical phonons ($\omega_{LO}/\omega_{TO} \rightarrow 0$), indicative of a strong coupling between vibronic and electronic degrees of freedom.

During the metavalent MIT the lattice, vibrational and electronic degrees of freedom are coupled to each other. This perspective helps to further understand the categorization of solids in table 1. In prototypical covalent and metavalent transition, the structural transition “S” (attainment of the high-symmetry, high coordination structure) coincides with the electronic transition “E”, defined by bandgap closure (S = E). The route to E differs between the two, however, metavalent compounds reach E through a continuous evolution of the electronic structure, while covalent solids reach the metallic state discontinuously. Indicative of the metavalent route is the participation of the vibrational modes which drive the structural change.

Increasing ionicity can decouple S and E such that the high coordination structure is reached while a finite band gap remains ($S \neq E$). This behavior can be described by the Phillips-Van Vechten description of the band gap $E_g^2 = E_h^2 + C^2$. Where $E_h$ denotes the homopolar contribution and C denotes the heteropolar, or ionic, contribution, which is mainly determined by the cation-anion electronegativity difference. Structural change can suppress the homopolar contribution. Whereas the heteropolar contribution can preserve a residual band gap (Compare discussion of figure S11 and S12). The MIT then occurs only upon further compression by continuous band broadening and therefore exhibits an ionic-like MIT signature.

The dependency of the band gap to pressure helps to unravel the relationship between atomic rearrangement, material properties and the bonding that drives them. To test whether bonding can explain these unusual changes across the MIT, chemical bonding in all solids discussed so far has been analyzed with the tools of quantum chemistry.

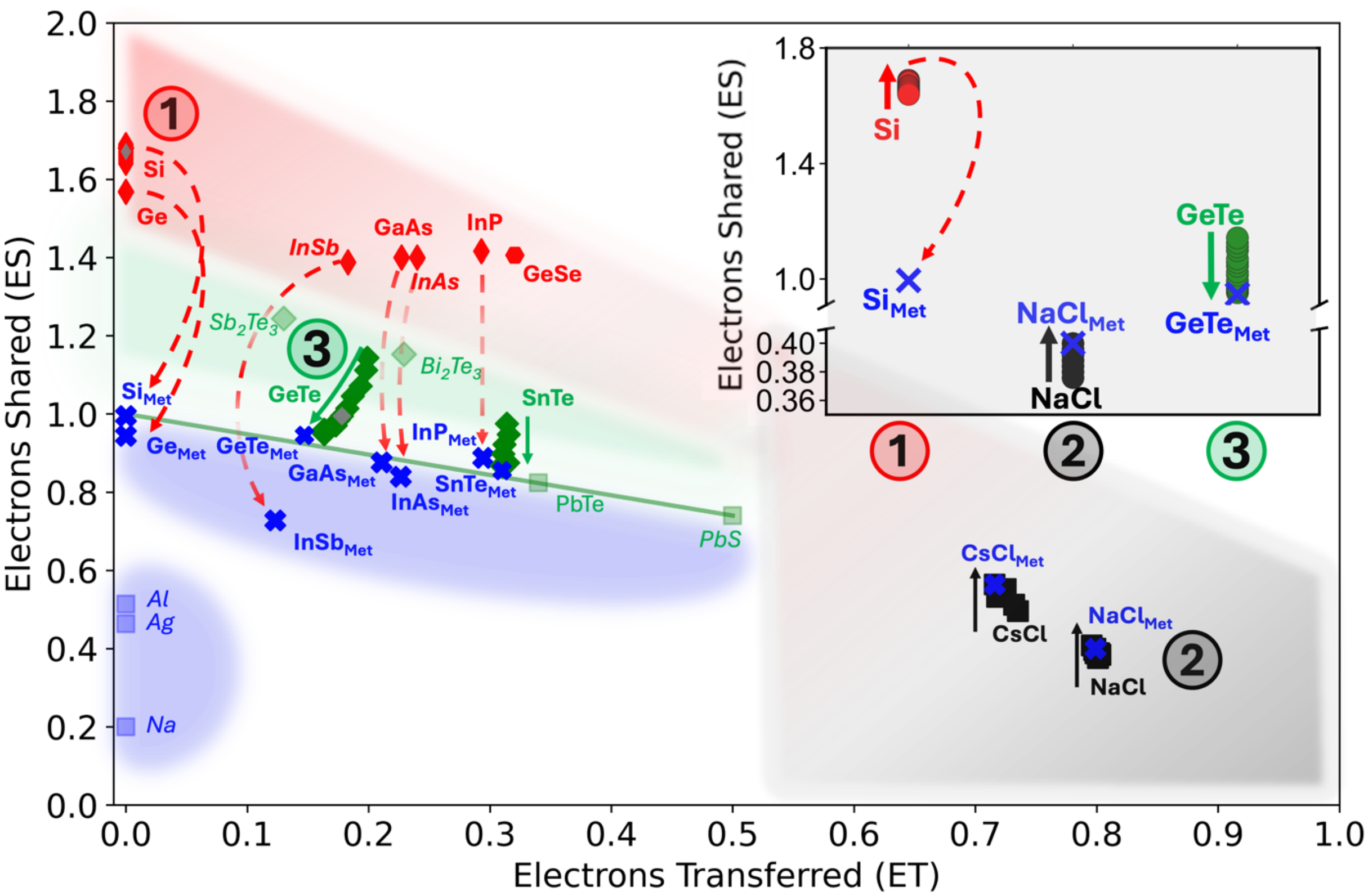

***Figure 4.*** *2D Map for chemical bonding in solids and pathways of pressure induced MITs for ionic, covalent and metavalent solids. The map is spanned by two quantum chemical bonding descriptors, the number of electrons shared between adjacent atoms and the number of electrons transferred (normalized by the oxidation state). The arrow shows how the different materials change their bonds upon increasing pressure, also schematically depicted in the inset. While ionic solids become metals at ES values much smaller than 1, for covalent and metavalent solids, the MIT proceeds at the border between metavalent and metallic bonding (green solid line). The MIT is discontinuous for covalent compounds, but continuous for metavalent and ionic ones.*

During the ionic MIT of NaCl ($Pm\bar{3}m$ structure), ES increases continuously but stays well below 1 (Figure 4 & Figure S13). Covalent and metavalent solids instead become metallic close to the green line, the border between metavalent and metallic bonding, where solids with perfect octahedral arrangement are found; it is characterized by small ET and ES ≈ 1, i.e., half an electron pair. Yet, covalent and metavalent solids reach this line via rather different routes.

Covalent solids like Si start at rather high ES values (1.57). Upon compression ES first rises slightly, as the orbital overlap increases, and then drops abruptly to about 1 at the first-order transition to the metallic β-Sn structure[16,17], where the sudden jump in the effective coordination number is accompanied by a concomitant decrease of ES. The same is observed for Ge and for the III–V semiconductors GaAs, InAs, AlAs, InSb, AlSb, InP and AlP (Figure S11 & Table S3).

Metavalent solids are the only ones whose ES decreases continuously during compression until they become metallic at the green–blue border, a trend closely related to the increasing effective coordination number as the Peierls distortion is suppressed[30]. Together with the distinct evolution of $Z^*_+$, $\varepsilon_\infty$, $\omega_{LO}$ and $\gamma_{LO}$ at the MIT, this uniquely distinguishes the metavalent MIT from covalent and ionic ones.

These bonding changes can now be linked to property changes – in particular the different behavior of $\omega_{LO}$ in metavalent and covalent solids. Figure 5a depicts the paths of Si and GeTe from the insulating to the metallic state. For the positions marked on the map, the potential energy surface (PES) has been calculated along the eigenvector of the (predominantly) longitudinal optical mode (Figure 5b), which in metavalent compounds corresponds to the ⟨111⟩ distortion, i.e., the relative shift of Ge and Te in the rhombohedral primitive cell.

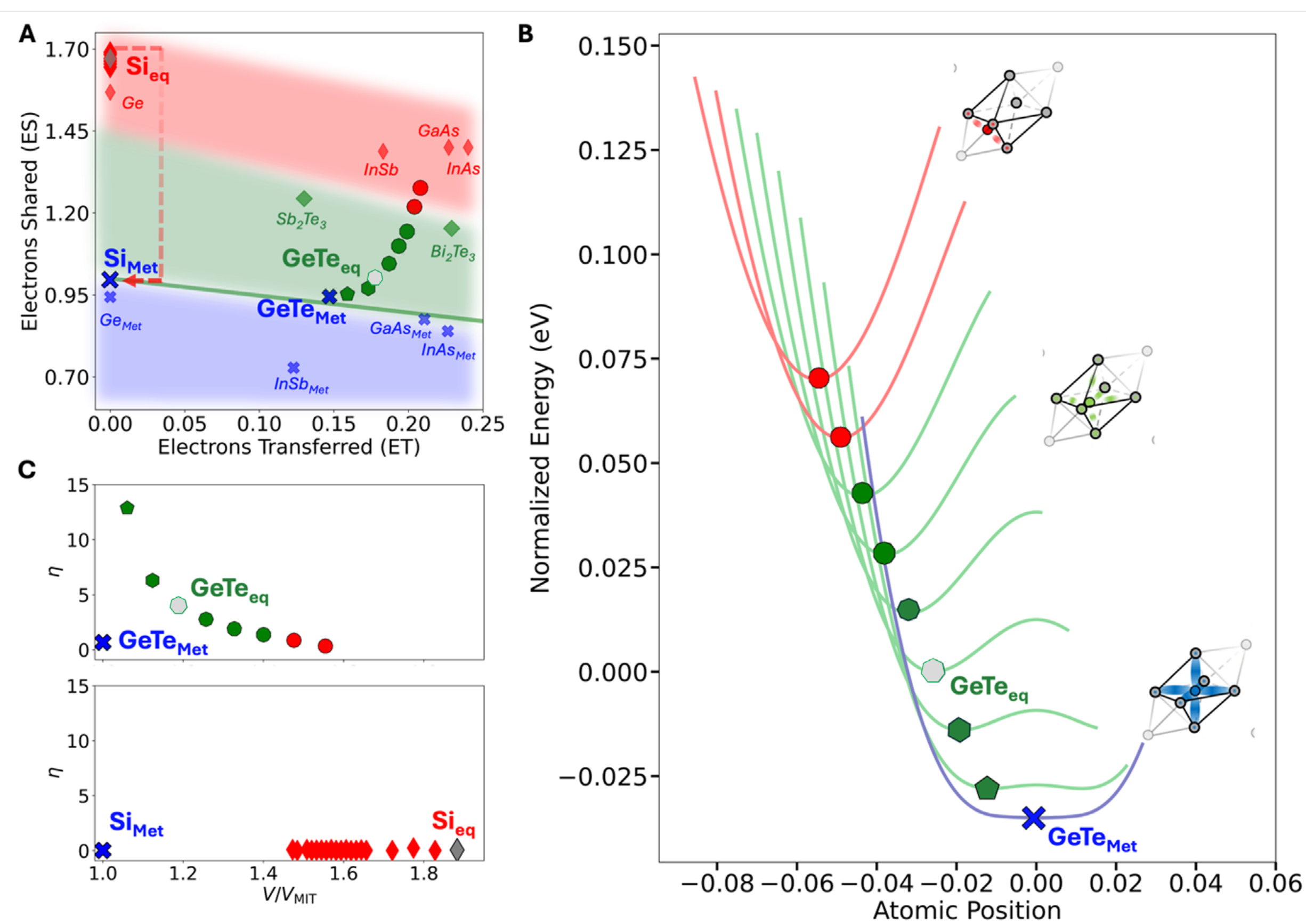

***Figure 5.*** *Potential energy landscapes close to the metal–insulator transition (MIT) for GeTe. Evolution of number of electrons shared and transferred on the path towards the metallic state for Si and GeTe (a). (b) Potential energy along the LO-mode of GeTe, showing an increasing anharmonicity, i.e., deviation from a harmonic potential with decreasing cell size. The energies are shifted with respect to the ground state of GeTe which is indicated by the subscript "eq". (c) parameter η quantifying the anharmonicity along evolution of the PES for GeTe and Si, see text for detailed description of η.*

The distortion is progressively suppressed with increasing pressure[31]. Initially the PES is deeply parabolic (a hard phonon); with decreasing cell size the two minima approach each other and the potential flattens markedly at the MIT.

Fitting a quartic function to the PES and taking the ratio η of the cubic to the quadratic term (Figure 5c) confirms this trend: approaching the MIT, the parabolic contribution decreases while the cubic term becomes increasingly dominant. For Si, in contrast, the harmonic behavior remains dominant up to its MIT. The approach to the metallic state in metavalent solids is thus closely linked to the degree of the Peierls distortion and its related properties; controlling this distortion therefore produces a distinct response to external stimuli, as observed in pump-probe experiments[32,33].

**Discussion:**

Three classes of pressure-induced MITs – ionic, covalent and metavalent – can thus be distinguished by the evolution of $E_g$, $\varepsilon_\infty$, $Z^*_+$, $\omega_{LO}$ and $\gamma_{LO}$, and they also differ in the concomitant changes of atomic arrangement and chemical bonding. Structural changes under pressure, such as the first-order transition of Si or GaAs, have been studied before and are fully reproduced here; the novelty lies in classifying the MITs by the systematic changes of properties and bonding. Remarkably, the three classes follow different routes on the bonding map: the ionic MIT involves no atomic rearrangement and follows a smooth path into the metallic state; the covalent MIT is a first-order transition in both structure and bonding; and the metavalent MIT is second order, with atomic arrangement and bonding changing gradually until the MIT is reached.

The most intriguing case is the metavalent MIT, where LO and TO modes soften simultaneously while $Z^*_+$ increases and $\varepsilon_\infty$ diverges. It is thus the only transition studied in which electronic and lattice degrees of freedom are strongly coupled – a coupling that justifies identifying metavalent solids as quantum materials. The hydrogen chain serves as a toy model for this MIT, sharing its characteristic features, including vanishing $\omega_{LO}$ and $\omega_{TO}$ and a very high $\gamma_{LO}$ at the MIT.

This resemblance clarifies the nature of metavalent bonding. The metallic hydrogen chain has one electron per atom but two equidistant neighbors, i.e., a half-filled band and a bond formed by just one electron – half an electron pair – in contrast to an ordinary covalent bond, which shares two electrons. This supports the interpretation of metavalent bonding as an electron-deficient interaction with approximately one electron shared between adjacent atoms, corresponding to a 2c–1e bonding picture[34,35].

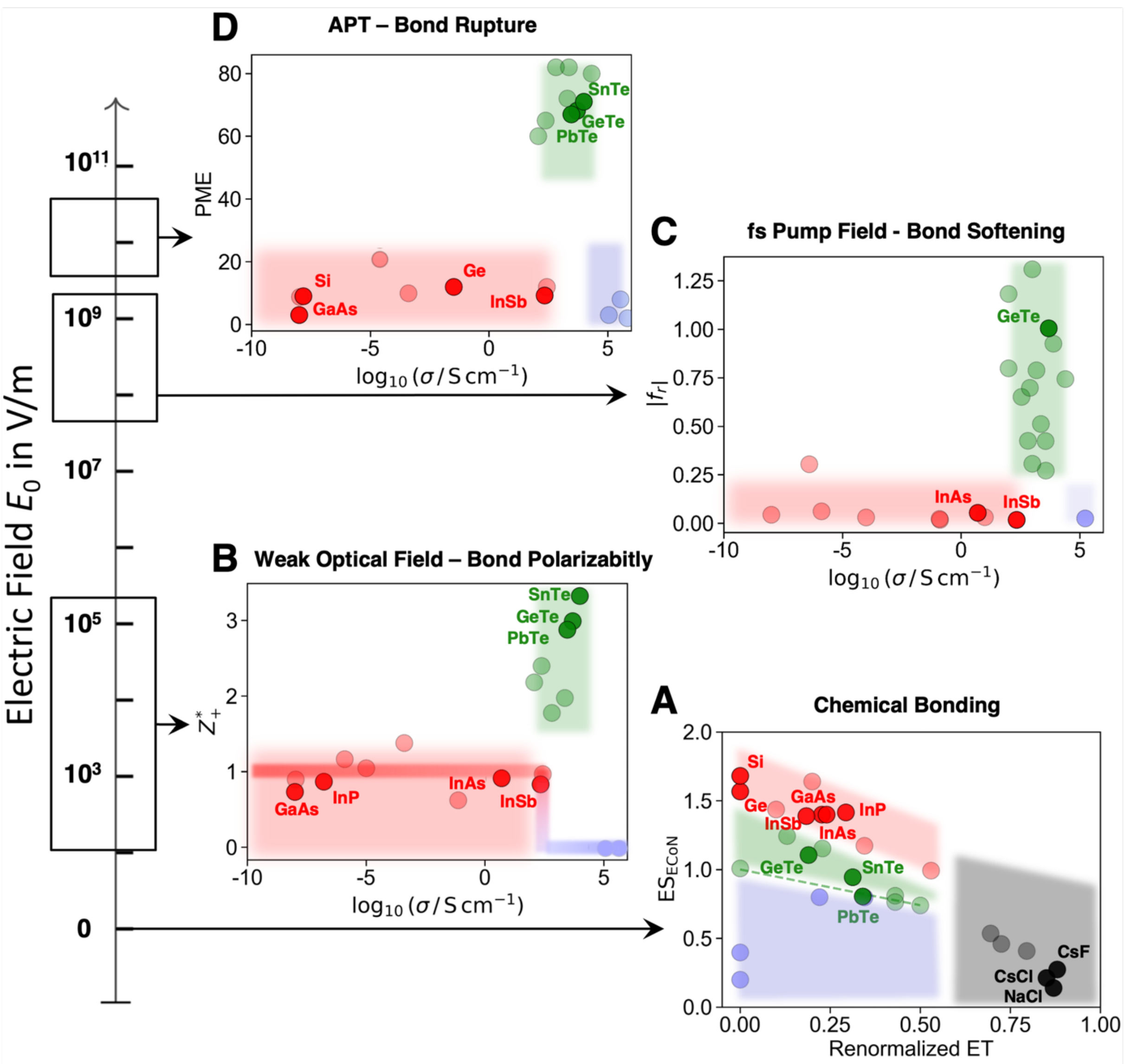

***Figure 6**: Different response functions for a broad range of applied electric fields up to $10^{10}$ V/m for different classes of solids. The ground state is described by a two-dimensional bonding map spanned by the number of electrons shared ES and the number of electrons transferred ET. The four colored background regions correspond to ionic (black), covalent (red), metallic (blue) and metavalent (green) bonding separated by a property-based material classification[27]. The solids which are classified here are labelled and marked by darker circles, while the lighter circles describe other solids for which data exist for different response functions. All data points shown here are also tabulated (Table S4) in the SI. B Three response functions upon increasing field strength are depicted, which show the same classification. Using external fields with a strength of $10^2$-$10^5$ V/m, the bond polarizability (Z*+) can be classified, C Bond softening, via the softening of phonon frequencies has been determined in the range of $10^8$-$10^9$ V/m D Finally, the bond rupture, characterized by the probability of multiple events (PME) as measured by APT is probed at fields around $10^{10}$ V/m.*

Since the electric dipolar field and the lattice distortions associated with optic phonons are the two ordering fields relevant to the MIT of metavalent solids, the responses to these fields at various orders exhibit anomalies that constitute the characteristic features of this MIT. Figure 6 A–D therefore relates chemical bonding to material properties by comparing response functions, i.e., the response of different solids to AC or DC electric fields of different strength.

Sorted by room-temperature conductivity, covalent solids typically lie below $10^2$ S/cm and metallic solids above $10^4$ S/cm; metavalent compounds fall in between, i.e., in the competition zone between electron localization and delocalization described by the Mott–Ioffe–Regel rule[36,37].

Figure 6 B–D compares three response functions sorted by increasing field strength. At $10^2$–$10^5$ V/m, the bond polarizability, i.e., the Born effective charge, is particularly high for metavalent solids (Figure 6 B); a related manifestation of their exceptional electric-field response is the anomalously large second-order Raman tensor, recently identified as a further fingerprint of metavalent bonding[38]. Metavalent solids are also characterized by a particularly pronounced bond softening upon intense laser excitation ($10^8$–$10^9$ V/m, Figure 6 C) and an unusual bond rupture in atom probe tomography (around $10^{10}$ V/m, Figure 6 D). That metallic, covalent and metavalent solids can be classified consistently by structure, bonding and response functions shows that characteristic differences in their wave functions govern both the ground-state properties and the response to external fields. Chemical bonding, i.e., the one- and two-electron densities, and material properties, i.e., the response functions, are thus intimately interwoven. This answers in the affirmative a question posed by Freeman Dyson [39] whether the wave functions of solids can be classified.

The present classification complements the Mott and Anderson pictures by addressing pressure-induced MITs in ordered crystalline solids, where ionic, covalent and metavalent materials follow fundamentally different mechanisms: continuous metallization with only minor lattice response, a first-order structural transformation to a higher-coordination phase, and – for metavalent solids – the simultaneous evolution of electronic structure, lattice dynamics and electric-field response, reflected in the anomalous $\varepsilon_\infty$, $Z^*_+$ and complete phonon softening identified here.

**Summary:**

Based on characteristic property changes, three classes of pressure-induced MITs have been identified. Five characteristic quantities were studied – the band gap, $\varepsilon_\infty$, the Born effective charge, the LO/TO frequencies at the Γ point and $\gamma_{LO}$ – which probe complementary aspects of the electronic, electron–lattice and lattice response. Ionic MITs proceed continuously through band-gap closure without significant change in atomic arrangement or electron–lattice coupling; covalent MITs occur via a first-order transition to a higher-coordination metallic phase; the most interesting, metavalent MIT combines a continuous change of atomic arrangement with a complete softening of the lattice and a diverging $\gamma_{LO}$, i.e., a simultaneous evolution of electronic structure, lattice dynamics and electron–lattice coupling. A 1D hydrogen chain with two atoms per unit cell serves as a toy model and shows that the metavalent MIT is a characteristic feature of electron-deficient solids at half filling.

For all three classes, ground-state properties of the wave function – the atomic arrangement and the one- and two-electron densities – are closely related to material properties, i.e., the response of the wave function to external perturbations such as electric fields. Wave functions of solids can thus be classified via their properties and their pressure-induced MITs, and quantum-chemical bonding descriptors provide a direct link between electronic wave function, atomic structure and experimentally accessible response functions. This indicates a close and fundamental relationship between chemical bonding and band structure, whose details remain to be unraveled; extending the classification to correlation-driven MITs is a natural next step.

**Acknowledgments:** The authors used an AI-assisted language tool for grammar and readability improvements. All scientific content and conclusions remain the responsibility of the authors.

**Funding:** The authors gratefully acknowledge funding by the German Science foundation within SFB917 (Nanoswitches) as well as the computing time provided to them at the NHR Center NHR4CES at RWTH Aachen University (project number p0022819).

**Author contributions:**

Conceptualization: MW

Methodology: MW, CFS, TB

Investigation: TB, CFS, RA

Visualization: TB, DK

Funding acquisition: MW

Supervision: MW, UW

Writing – original draft: TB, MW

Writing – review & editing: MW, TB, UW, RA, CFS, DK

**Competing interests:** The authors declare that they have no competing interests

**Data, code, and materials availability:** The data that support the findings of this study are available from the corresponding author upon reasonable request.

# Supplementary Materials

## Materials and Methods:

Density functional theory (DFT) calculations were performed with Quantum ESPRESSO 7.4[40], using PAW pseudopotentials (Dal Corso) and the PBE exchange–correlation functional (GGA). Total energies and Kohn–Sham band gaps ($E_g$) were obtained from self-consistent pw.x calculations; Γ-point phonon frequencies ($\omega_i$), the high-frequency dielectric tensor ($\varepsilon_\infty$) and Born effective charge tensors ($Z^*$) were computed with ph.x within density-functional perturbation theory (DFPT)[41].

Prior to the property calculations, the Peierls-distorted systems (GeTe, SnTe, PbTe, GeSe) were relaxed at a series of fixed unit-cell volumes by shifting the central atom along ⟨111⟩, yielding a sequence of volumes with different degrees of Peierls distortion; the rhombohedral angle was fixed at 60° to save the cost of an additional cell degree of freedom. For zincblende III–V and group-IV compounds, high-pressure configurations were generated by hydrostatic volume reduction at ideal zincblende coordinates, which have no internal degrees of freedom.

High-pressure structures of ionic and covalent compounds were relaxed with respect to cell volume/shape and atomic positions at fixed target pressures (vc-relax) until the total energy changed by less than $10^{-10}$ between ionic steps and all force components were below $10^{-5}$ Ry/bohr. All ground-state SCF calculations (GeTe, SnTe, PbTe, GeSe, GaAs, Si, Ge, NaCl, CsCl, CsF, AlAs, AlN, AlP, AlSb, GaN, InAs, InN, InP, InSb) used at least a 15 × 15 × 15 k-point mesh – for non-cubic materials a mesh with at least 1000 k-points – wavefunction and charge-density cutoffs of 100 and 400 Ry, and an electronic convergence threshold of $10^{-10}$ Ry; for metals and small-gap semiconductors Gaussian smearing of 0.001–0.005 Ry was applied. Phase stabilities were determined by zero-temperature enthalpy comparison ($\Delta H = \Delta U + p\Delta V$) with the same parameters.

To verify that the different structure-specific relaxation procedures did not introduce systematic differences in the calculated structural evolution, additional calculations were performed for representative metavalent, ionic, and covalent compounds, namely GeTe, NaCl, and GaAs, respectively. For this comparison, the same unconstrained variable-cell relaxation procedure was applied to all three systems using vc-relax with cell_dofree = 'all', allowing both the cell shape and volume to relax without additional constraints. The resulting structures and their evolution under compression were consistent with those obtained using the structure-specific relaxation procedures (compare

Figure S22). Consequently, the relaxation protocols adapted to the individual material classes were employed throughout the study to reduce the computational cost.

Phonon frequencies were calculated within DFPT (ph.x), starting from the self-consistent ground state obtained with a 40 × 40 × 40 k-point grid for GeTe, SnTe, PbTe, GeSe, GaAs, Si, Ge, NaCl, CsCl and CsF. The dynamical matrix, built from the second derivatives of the total energy with respect to atomic displacements, was diagonalized to obtain frequencies and eigenvectors; at Γ, dielectric tensors and Born-effective charges were evaluated within DFPT where possible. A phonon convergence threshold of $10^{-14}$ was used, without smearing for insulators and with Gaussian smearing of 0.001–0.005 Ry for metals. The mode-specific Grüneisen parameter was obtained as

$$\gamma_{\mathrm{i}} = -\frac{V}{\omega}\frac{\partial\omega_{\mathrm{i}}}{\partial \mathrm{V}}.$$

For the one-dimensional hydrogen chain, a k point grind of 100x1x1 k-points was used and the chain was constructed along the x direction of the unit cell. The other direction was used as vacuum layers with ~ 20 Å vacuum thickness between individual chains. A kinetic energy cutoff of 100 Ry and a charge density cutoff of 400 Ry. The metallic structures were calculated using 0.001 Ry smearing width.

In addition, ground-state Kohn–Sham wavefunctions of all structures were obtained with ABINIT (PBE-GGA, PAW potentials, 10 × 10 × 10 k-point grid, plane-wave cutoff 12.5–17.5 Ha; 0.005 Ha smearing for metals, all calculations being run with and without smearing) and analyzed with DGrid 4.7 to obtain Quantum Theory of Atoms in Molecules (QTAIM) quantities. DGrid constructs a real-space charge-density grid from the wavefunction, determines Bader basins and integrates the exchange–correlation density over both electron coordinates within one basin (localization index) or over two different basins (delocalization index, DI). Subtracting the nominal charge of the free reference atom yields the total number of electrons transferred (TET = $N_a - Z_a$), and division by the formal oxidation state the number of electrons transferred (ET = TET/|Ox|).

The effective coordination number (ECoN) was calculated following Hoppe[26]. In contrast to nearest-neighbor schemes, it is a distance-weighted measure of coordination that assigns higher weights to closer neighbors,

$$ECoN = \sum_j exp\left[1 - \left(\frac{d_j}{d_r}\right)^6\right]$$

], where $d_j$ is the distance to neighbor j and the effective distance $d_r$ is given by

$$d_r = \frac{\sum_j d_j \ \exp\left[1 - \left(\frac{d_j}{d_1}\right)^6\right]}{\sum_j \exp\left[1 - \left(\frac{d_j}{d_1}\right)^6\right]}$$

# Additional Figures and Text

The following paragraphs should support the findings of the main manuscript and are presented in chronological order in which they appear in the main manuscript. Therefore, the property evolution of Ionic/Covalent and Metavalent materials are discussed first. In addition to the materials presented in the main part additional materials are presented. The discussion of properties is followed by the discussion of the structure and bonding. Additionally, a systematic analysis of GeSe is conducted. The section is closed by a summary of the classification where references and data created in the framework of this work are presented together.

Table of Contents

# 1. Properties Ionic/Covalent/Metavalent Solids:

## 1.1. Properties of Ionic Compounds:.

The evolution of properties for the ionic systems is presented in Figure S1. CsF follows the same property trend observed for NaCl during the metal–insulator transition. Both transitions show a continuous closure of the bandgap (Figure S1 A, F), an increase of $\varepsilon_\infty$ (Figure S1 B, G) and a nearly constant behavior of $Z_+^*$ (Figure S1 C, H). The evolution of the phonon modes is shown in Figure S1 D and I. The mode-specific Grüneisen parameters were obtained by fitting the phonon frequencies with third-order polynomials and are shown in Figure S1 E–J.

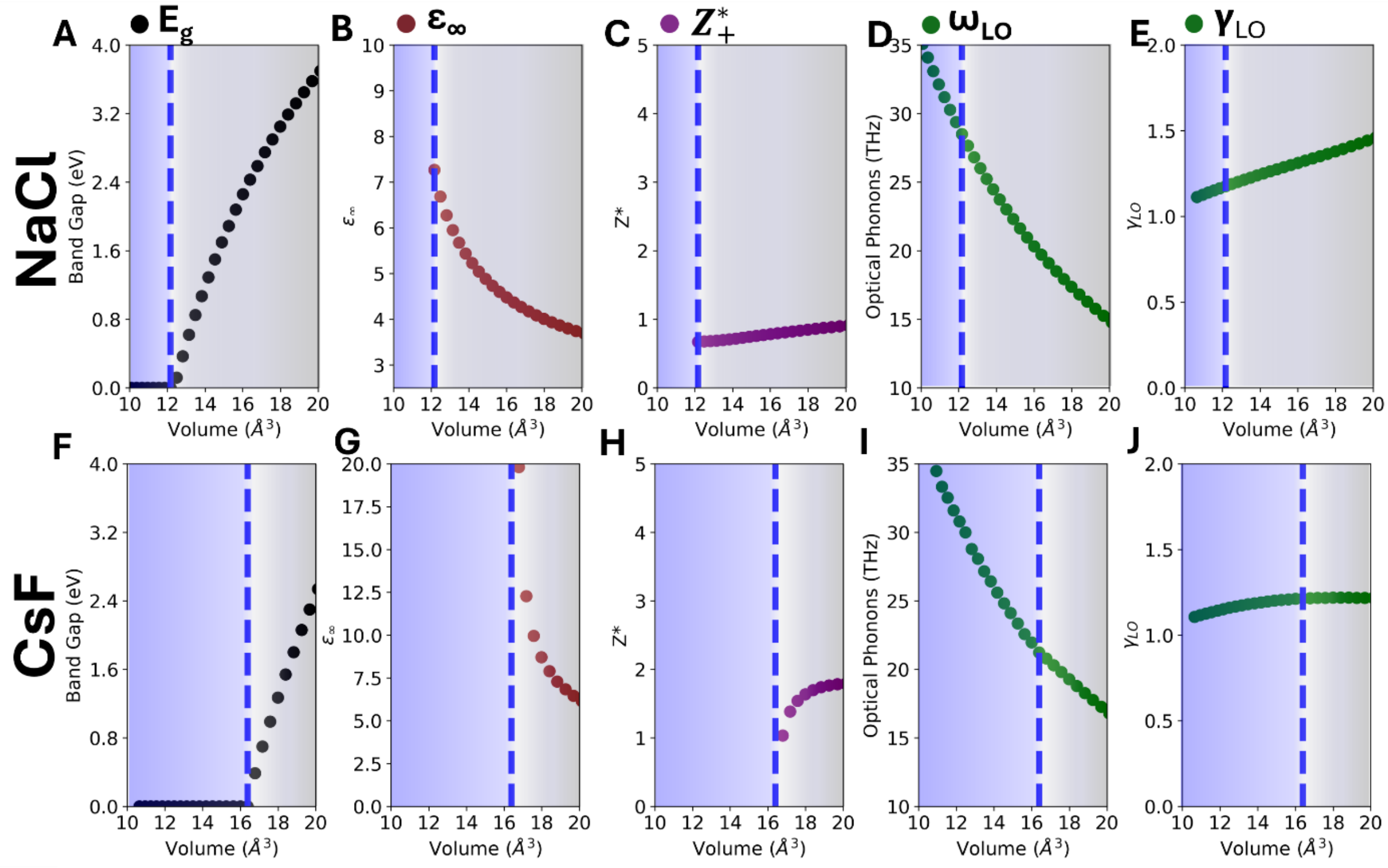

***Figure S1:*** *Properties of NaCl (A–E) and CsF (F–J) in the $Pm\bar{3}m$ structure. The materials exhibit an evolution of properties that is categorized as an ionic metal–insulator transition. The transition to the metallic state proceeds continuously, as indicated by the gradual closure of the band gap (A, F) and the divergence of $\varepsilon_\infty$ (B, G)). The Born effective charge (C, H) remains low throughout the MIT, indicating weak electron–phonon coupling. The phonon modes (D, I) do not provide a clear indication of the transition to the metallic state, and no significant increase in anharmonicity is detected during the transition. The mode-specific Grüneisen parameters (E, J) for NaCl and CsF were evaluated by fitting third-order polynomials to the volume dependence of the phonon modes.*

NaCl and CsF were studied in the $Pm\bar{3}m$ phase, which is known to represent a high-pressure phase of several ionic compounds. Pressure-induced metallization within this structure has been reported in the literature and can occur without an accompanying structural transition. Since typical alkali halides metallize only at comparably high pressures, for example, approximately 259 GPa for NaI[42], it is important to exclude the possibility that a thermodynamically competing phase is already metallic at these pressures. Therefore, structures predicted to represent the thermodynamically stable high-pressure phases were also investigated.

The oC8 structure for NaCl has been predicted to represent the high-pressure phase for pressures ranging from 322 to 645 GPa[13]. The same study suggests a band gap closure at approximately 584 GPa. To quantify the evolution of properties, $E_g$, $\varepsilon_\infty$, $Z_+^*$ were calculated in the range from 500 to 620 GPa, as shown in Figure S2 (A–C). The progression exhibits clear similarities to that observed for the $Pm\bar{3}m$ phases. Additionally, the Pbam phase was investigated for CsCl (Figure S2 (D–F)), which has been reported to be the high-pressure structure[43]. Like the high-pressure phase of NaCl, the Pbam structure exhibits analogous behavior of properties under pressurization to the $Pm\bar{3}m$ structure, as visualized in Figure S2.

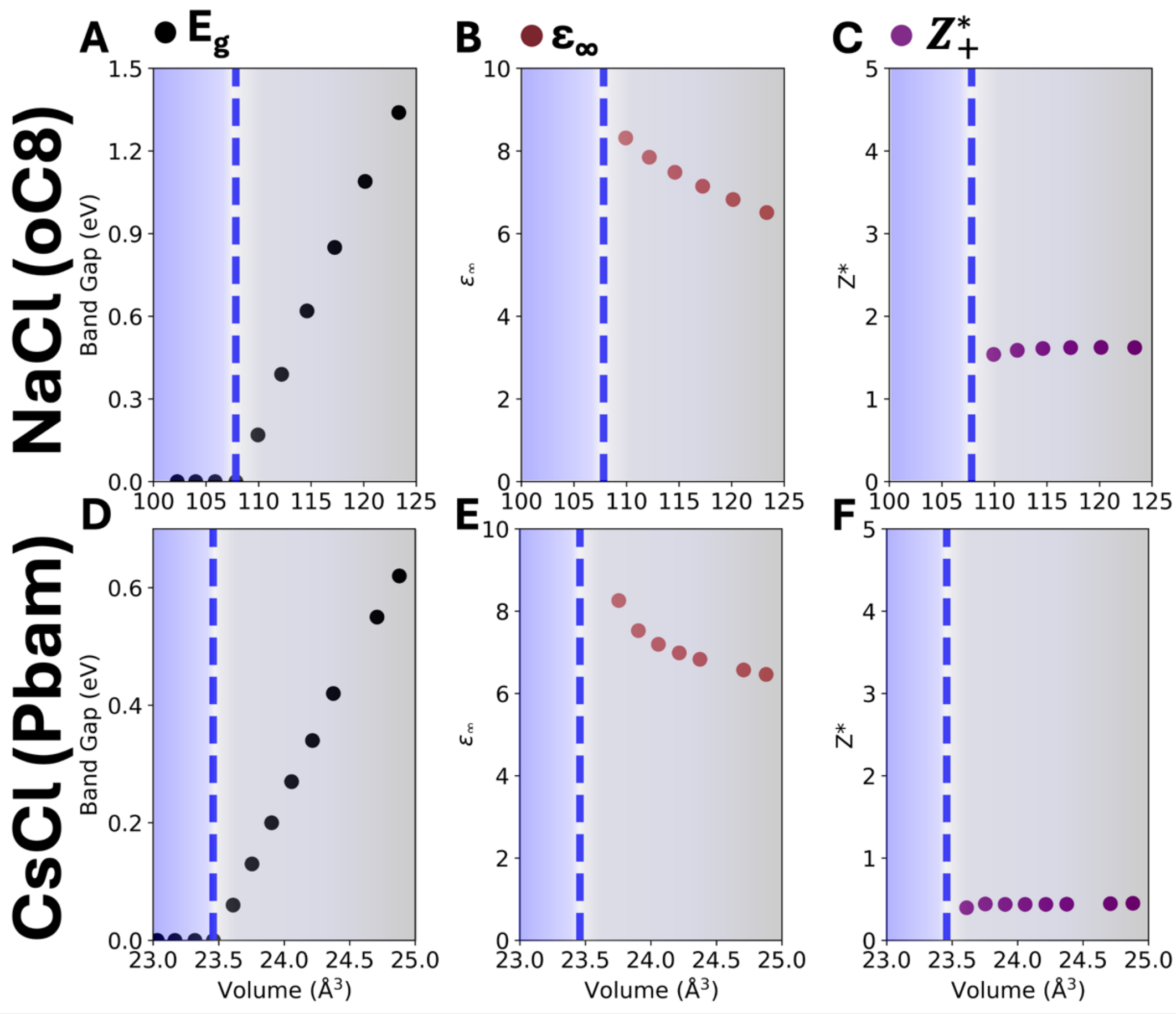

***Figure S2:** Evolution of $E_g$, $\varepsilon_\infty$, $Z_+^*$ for the oC8 and Pbam phase of NaCl (A-C) and CsCl (D-F). The structures represent the predicted high-pressure phases for both compounds. In comparison to the progression of properties in the $Pm\bar{3}m$ no significant difference can be observed.*

Beyond these ionic materials, studies report on the metallization of individual ionic systems[42-44] for example, experimental metallization of CsI was achieved by Prof. Eremets using a diamond anvil cell, where it was noted that the metallization mechanism is insensitive to the crystal structure of CsI, supporting the notion of a characteristic MIT for ionic materials. It has also been shown for BaS, BaSe, BaTe, NaI, KI and RbI that these compounds undergo a continuous electronic transition to the metallic state without structural participation[42,44,45]. These materials have been classified as ionic compounds[27] and are also consistent with our definition of an Ionic MIT

# *Properties of Covalent Compounds:*

The covalent compounds investigated in this work include GaAs, Si and Ge. The evolution of *in* $E_g$, $\varepsilon_\infty$, $Z_+^*$, $\omega_{LO}$, $\gamma_{LO}$ can be seen in Figure S3 (A-O).

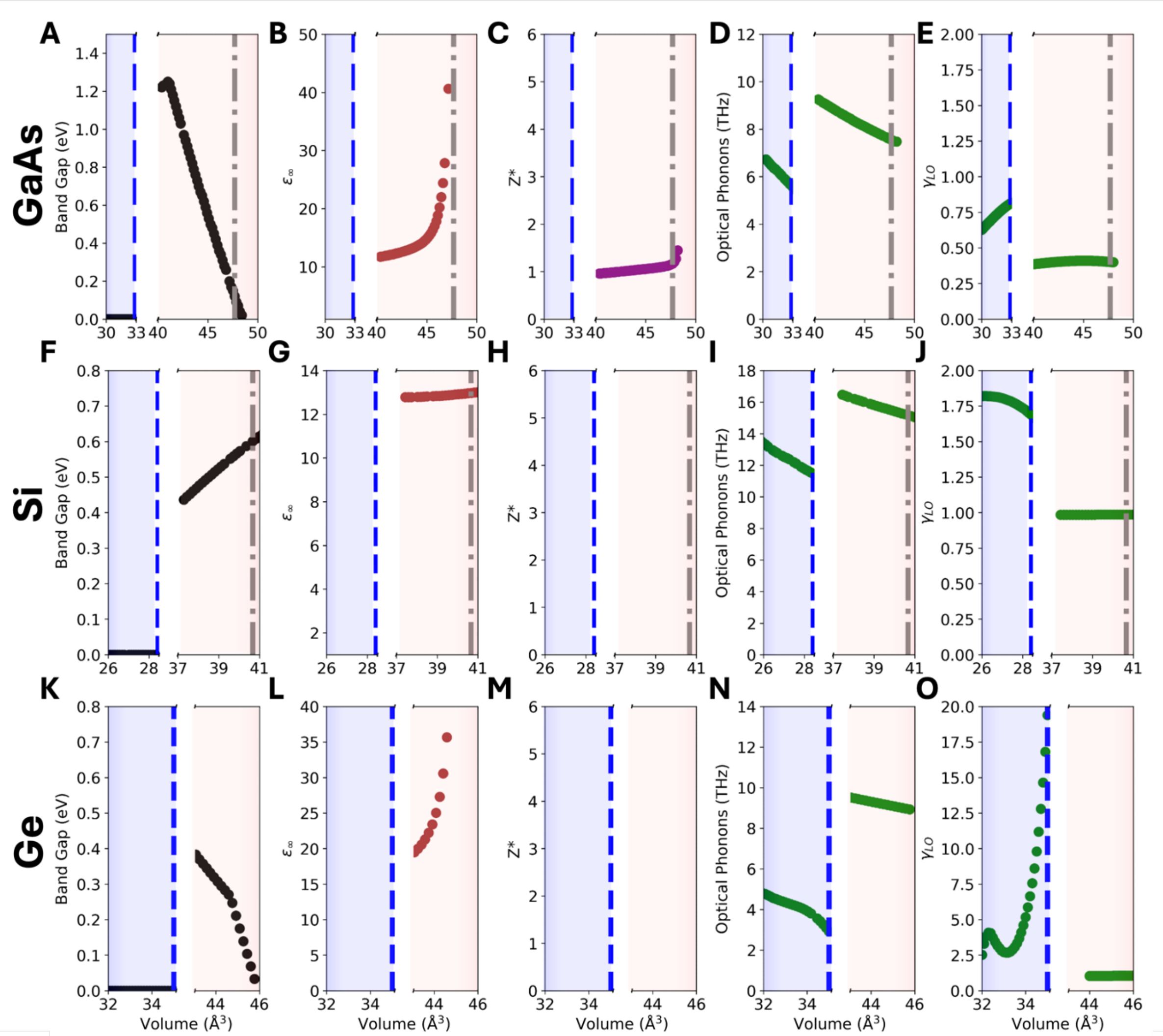

***Figure S3:*** *Progression of $E_g$, $\varepsilon_\infty$, $Z_+^*$, $\omega_{LO}$, $\gamma_{LO}$ for GaAs (A-E), Si (F-J), and Ge (K-P). The materials show an evolution of properties that is categorized as a covalent metal–insulator transition. In all three cases, the transition to the metallic state occurs discontinuously, as indicated by the spontaneous closure of the band gap.*

The change in the slope of $E_g(V)$ in Figure S3 A,F,K is related to the behavior of the highest occupied state and lowest unoccupied state in the 1st Brillouin zone of each individual compound e.g the direct Γ-Γ bandgap (Figure S4 A) of GaAs increases with pressure while the Γ-X gap (Figure S4 B) has a negative dependence on pressure. As a result, a direct to indirect transition occurs at a volume of approximately 40.667 Å$^3$in GaAs leading to the change of

${dE_g(V)}/{dV}$ in Figure S6 A.

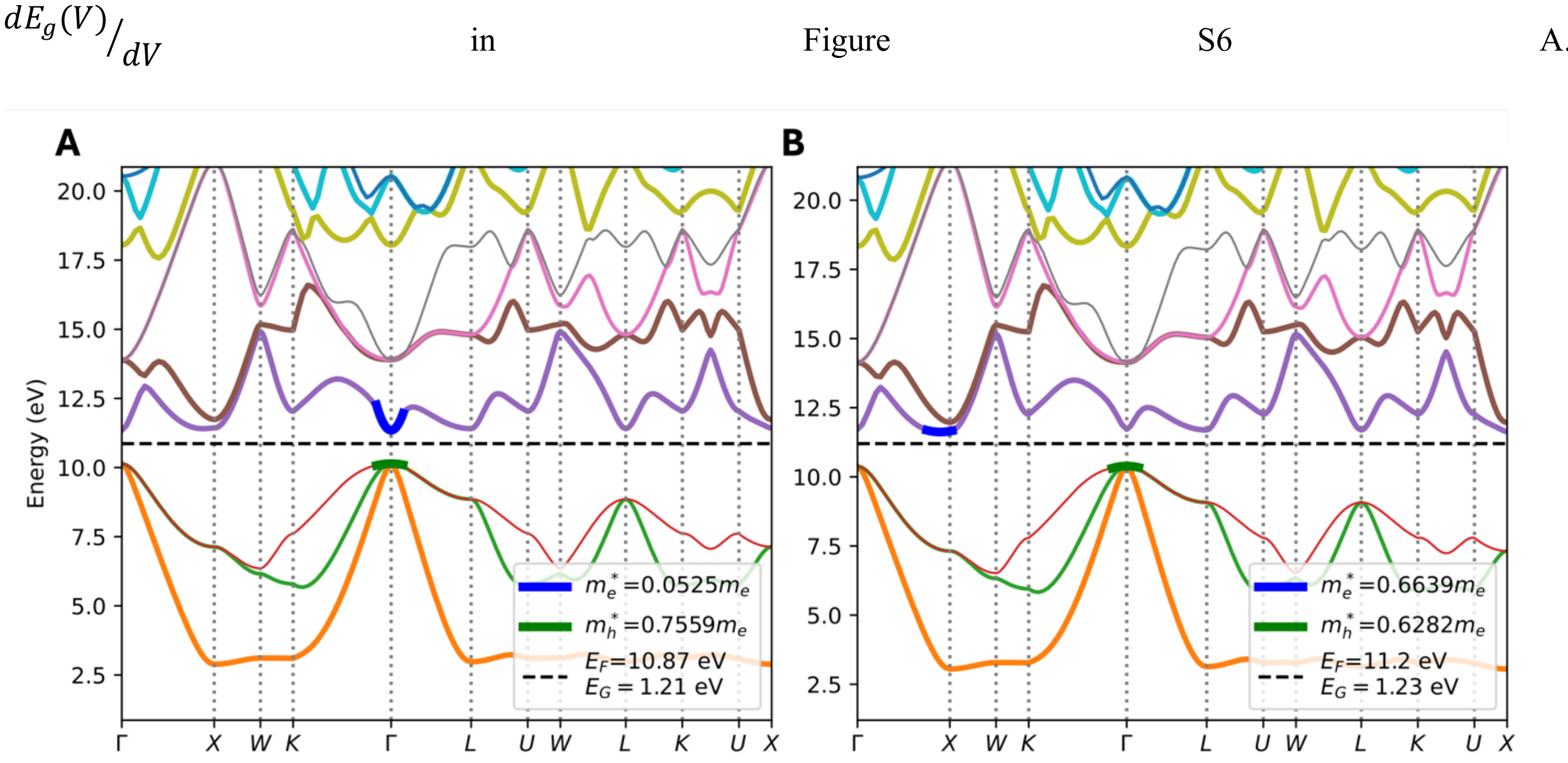

***Figure S4:** band structures of GaAs at volumes around 40.667 $Å^3$. A change from a direct bandgap (A) to an indirect bandgap (B) can be observed which is related to different dependencies on pressure for these points in the 1st Brillouin zone.*

### *1.1.1. Additional Covalent Compounds:*

Besides the three compounds investigated the first order phase transition has also been calculated for several other members of the III-V compounds and the high-pressure phase has been checked for metallic behavior. To determine the high pressure structure, data from literature was used and the transition has been calculated based on the structure reported in literature[14,16]. The transition pressures, high-pressure structures, and whether the transitions are associated with an MIT are summarized in Table S1. The metallic state after the first order phase transition has been compared to literature to ensure that the metallic state is not an artifact of the functional used.

***Table S1:** All investigated III–V compounds and their corresponding ground-state structures are shown. In addition, the high-pressure structures and corresponding transition pressures, calculated from enthalpy differences, are provided. If a compound*

*exhibits a metal-to-insulator transition, an "×" is shown in the MIT column. The thermodynamic stable high-pressure structures were taken from the literature[14].*

| Compound | Eq. Structure | High Pressure Structure | MIT | $P_{MIT}$ [GPa] |
|---|---|---|---|---|
| | | | | |
| GaAs | $Fd\bar{3}m$ | $Fm\bar{3}m$ | x | 14.55 |
| Si | $Fd\bar{3}m$ | $I4_1/amd$ | x | 9.37 |
| Ge | $Fd\bar{3}m$ | $I4_1/amd$ | x | 8.34 |
| AlSb | $Fd\bar{3}m$ | $Cmcm$ | x | 5.40 |
| AlAs | $Fd\bar{3}m$ | $P6_3/mmc$ | x | 8.60 |
| AlP | $Fd\bar{3}m$ | $P6_3/mmc$ | x | 9.35 |
| AlN | $P6_3mc$ | $Fm\bar{3}m$ | | 12.79 |
| InSb | $Fd\bar{3}m$ | $Cmcm$ | x | 3.32 |
| InAs | $Fd\bar{3}m$ | $Fm\bar{3}m$ | x | 3.62 |
| InP | $Fd\bar{3}m$ | $Fm\bar{3}m$ | x | 8.48 |
| InN | $P6_3mc$ | $Fm\bar{3}m$ | | 12.53 |
| GaN | $P6_3mc$ | $Fm\bar{3}m$ | | 45.01 |

### *1.1.2. Bandgap discussion of Covalent Solids:*

DFT is known to underestimate the magnitude of the fundamental band gap. This issue is particularly evident for Ge, for which standard DFT predicts a vanishing or even negative band gap at the equilibrium volume. Rather than relying on the absolute band-gap value obtained from a single-point calculation, we therefore focus on its evolution over a sequence of different volumes. Although the absolute band gap is underestimated, its volume and pressure dependence is generally reproduced more reliably. Consequently, the calculated trend can remain meaningful even when the absolute values are systematically shifted relative to experiment. In the following table, the deformation potentials of GaAs, Si, and Ge obtained in this work are compared with experimental and theoretical values reported in the literature.

***Table S2:*** *Evaluated pressure dependences of the band gap for GaAs, Si, and Ge in meV/kbar along different directions in the first Brillouin zone, as indicated in the "Gap" column. Since some literature sources do not clearly specify whether the reported pressure dependence is linear or quadratic, both values are shown in the second column. The first value in the second column is obtained*

*from a linear fit, while the second value corresponds to the derivative of a quadratic fit evaluated at P=0. The literature values in the fourth and fifth columns are taken from[46].*

| Compound | Gap | Deformation Potential [meV/Kbar] | $Literature_{Exp.}$ [meV/Kbar] | $Literature_{Theo.}$ [meV/Kbar] |
|---|---|---|---|---|
| **Si** | $\Delta E_X$ | -1.79<br>-1.90 | - | -1.90 |
| **Ge** | $\Delta E_L$ | 3.70<br>4.80 | - | 4,34 |
| | $\Delta E_\Gamma$ | 12.2<br>14.3 | - | 12.9 |
| **GaAs** | $\Delta E_\Gamma$ | 9.38<br>11.5 | 8.5-12.6 | 9.8 |
| | $\Delta E_X$ | -1.8<br>-2.2 | - | -2.44 |

Besides the band gap, $\varepsilon_\infty$ was also calculated. All three investigated covalent compounds follow the Moss relation, given by the following equation. This comparison serves as a consistency check to assess whether the evolution of the DFPT-calculated $\varepsilon_\infty$ with compression is physically reasonable.

$$\varepsilon_\infty^2 * E_g = 95 \quad (1)$$

Figure S5 A-C depicts the evolution of $\varepsilon_\infty$ calculated using DFPT and estimated using the Moss rule. Both approaches show a divergence of $\varepsilon_\infty$ close to the equilibrium structure (large volumes) of GaAs (Figure S3 A) and Ge (Figure S3 C), which is related to the underestimation of the band gaps of these compounds. This effect is particularly pronounced for Ge, which is metallic in its DFT ground state. Upon compression, the band gap increases. The overall evolution under compression is captured well, with both curves exhibiting similar trends. Most importantly, close to the MIT, the behavior predicted by each method converges, and neither approach shows any indication of an impending structural & electronic transition, which is a characteristic of the covalent MIT close to the MIT the behavior of both methods approaches each other, and no indication of an upcoming phase transition can be observed, which is a central characteristic of the covalent MIT.

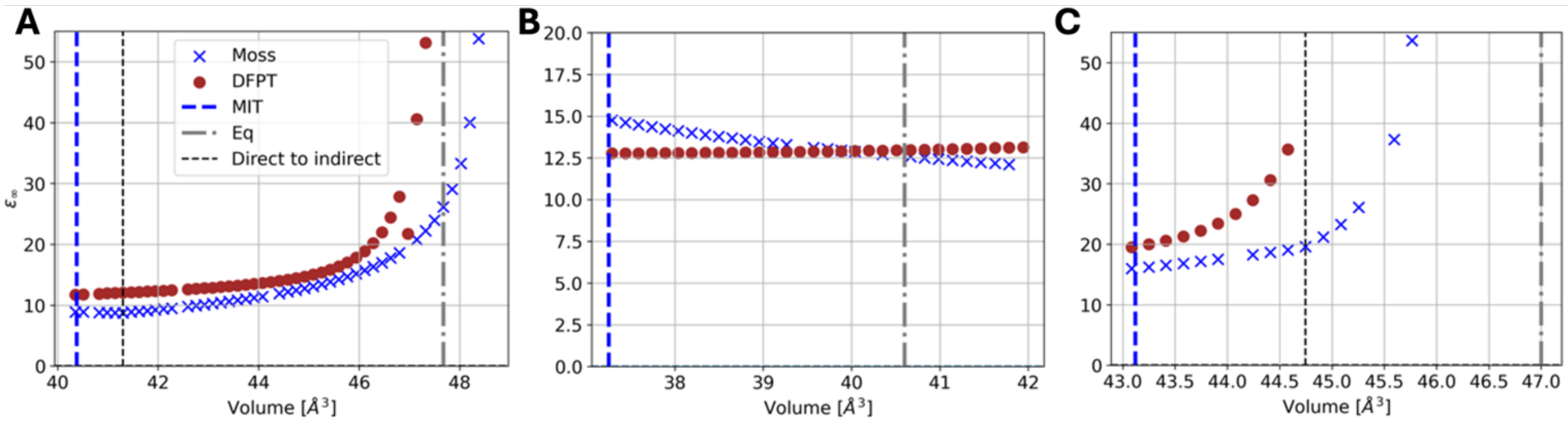

***Figure S5:*** *Comparison of ε∞ calculated using DFPT and the Moss relation given in Equation (1). In addition, the position of the equilibrium structure is indicated by a vertical grey dashed line, while the position of the MIT is indicated by a vertical blue line. The band gaps of GaAs and Ge are significantly underestimated at their equilibrium structures, leading to unusually high values of ε∞-. Since the evolution of the band gap under compression is captured well, the corresponding evolution of ε∞ toward the metallic state is also reproduced accurately.*

## *1.2. Properties of Metavalent Solids:*

Materials investigated in the framework of this study which exhibit Metavalent bonding (in the R3m phase) are GeTe, SnTe, PbTe and GeSe. Figure S6 A-T presents the evolution of $E_g$, $\varepsilon_\infty$, $Z^*_+$, $\omega_{LO}$, $\gamma_{LO}$ for all calculated systems. GeTe was already described in the main manuscript and a clear correspondence to the progression of the other compounds can be observed. The progression of the electronic identifiers in SnTe and GeSe show similarities to the progression observed in GeTe. For both system epsilon infinity depicts a significant increase when approaching the MIT (Figure S6 B, G, L, Q). Furthermore, a significant increase of $Z^*_+$ becomes apparent for all metavalent when approaching the MIT. This can be seen as an indicator of increased coupling of electronic and vibronic states, especially the zone center phonon mode which drives the distortion and opens a bandgap at the zone boundary of the 1st Brillouin zone (Figure S6 B, G, L, Q).

The vibrational frequencies $\omega_{LO}$ & $\omega_{TO}$ (Figure S6 D, I, N, S) decrease significantly when approaching the MIT. The simultaneous closure of the bandgap in addition to the softening of the phonons is the characteristic of the metavalent MIT.

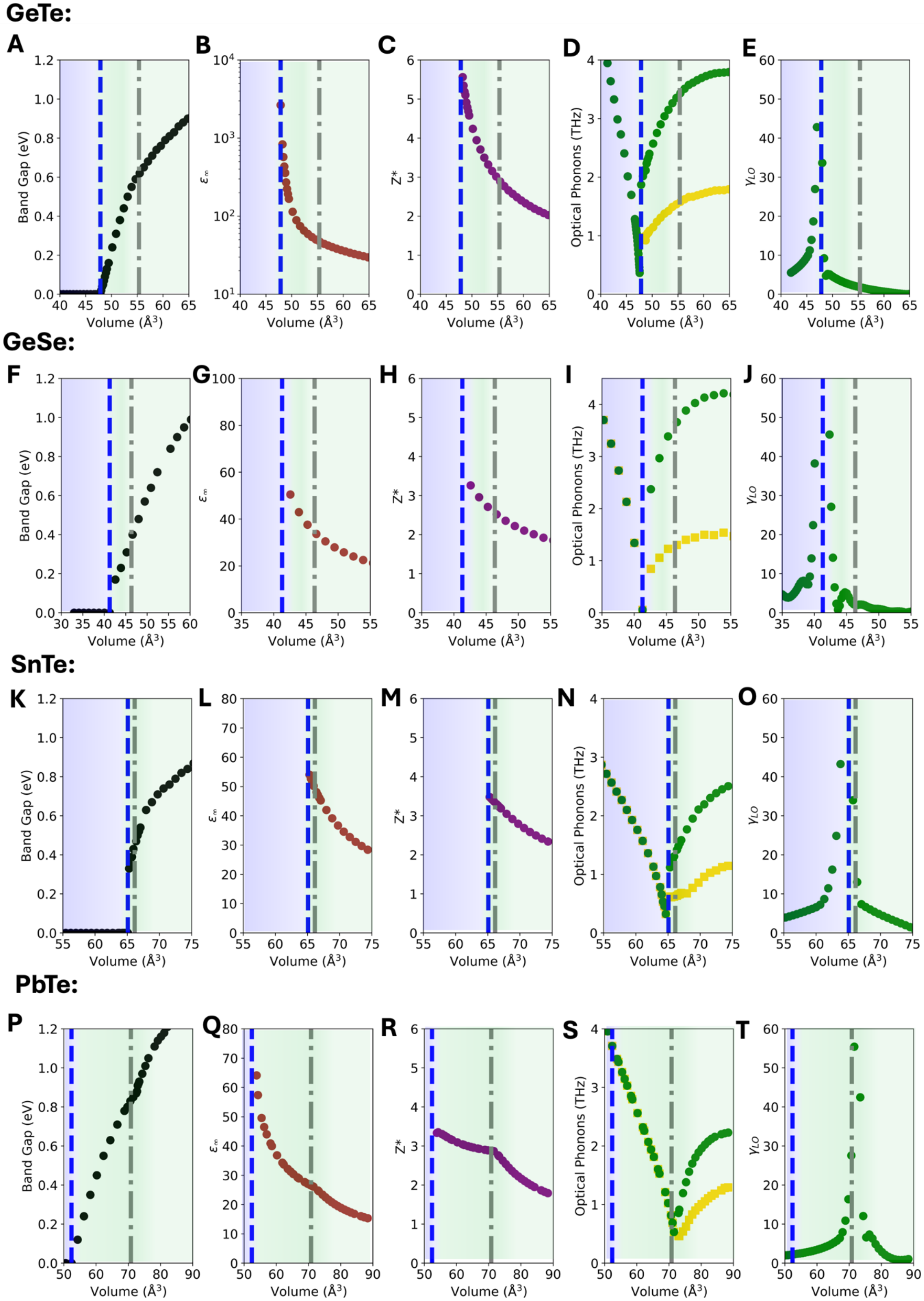
GeTe:
A
B
C
D
E
Band Gap (eV)
Optical Phonons (THz)
Volume (Å³)
GeSe:
F
G
H
I
J
SnTe:
K
L
M
N
O
PbTe:
P
Q
R
S
T

***Figure S6:*** *Properties of GeTe (A-E), GeSe (F-J), SnTe (K-O) and PbTe (P-T). For all compounds the Band Gap, $\varepsilon_\infty$, Z*, $\omega_{LO}$, $\gamma_{LO}$ are shown over the full range of compression. The vertical dashed blue line indicates the metal-to-insulator transition. The vertical dashed-dotted grey line indicates the equilibrium structure of the compounds.*

# 2. Structure:

## 2.1. Structure of Ionic Solids

The structural evolution of NaCl, CsCl, and CsF in the $Pm\overline{3}m$ phase is depicted in Figure S7. Here, it is evident that the structure remains unchanged during compression, as shown in Figure S7 (A, D, G). The effective coordination number (ECoN) remains constant at 11.47. Meanwhile, the number of shared electrons increases systematically with compression (Figure S7 C, F, I). By contrast, the number of transferred electrons exhibits significant scatter in the $Pm\bar{3}m$ phase.

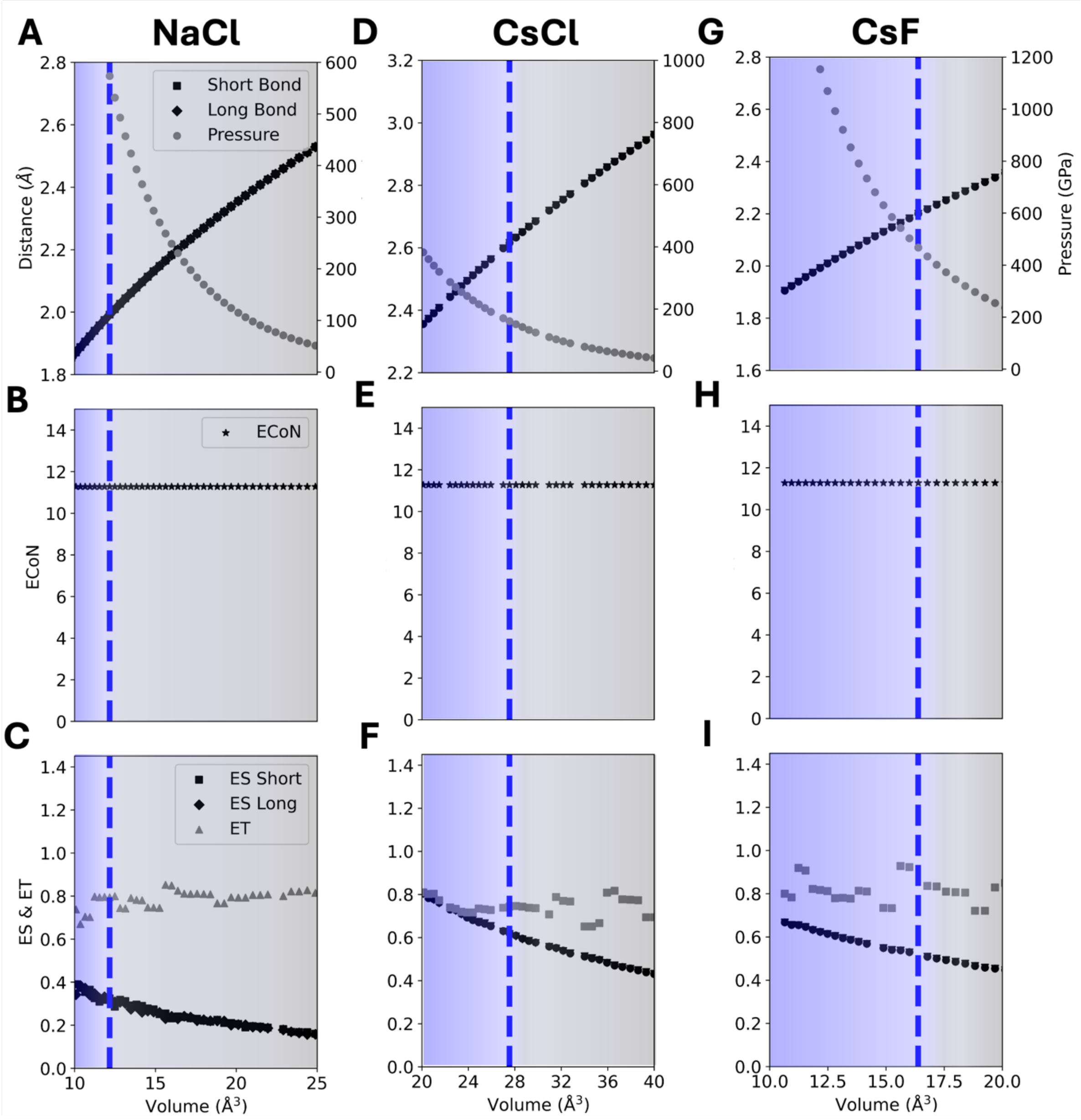

*Figure S7: Evolution of structure, chemical bonding and properties for NaCl (A-C), CsCl (D-F), and CsF (G-I). To characterize the structural progression the short and long bond as well as the effective coordination number (ECoN) is depicted.*

The metallization of NaCl and other ionic compounds was studied in the $Pm\bar{3}m$ phase, where our results reproduce findings reported in the literature[47]. Similar observations were made for a variety of alkali halides under pressure, which were compressed in the $Pm\bar{3}m$ phase until metallization[42] , induced by a continuous band gap closure. To ensure that the observations made are not an artifact of investigating the $Pm\bar{3}m$ structure alone, the predicted high-pressure structures (oC8 for NaCl and Pbam for CsCl) were examined, within the predicted pressure range reported in literature. Figure S8 (A–F) depicts the structure and bonding quantifiers for these high-pressure phases. Similar observations can be drawn to those observed for the discussion of the $Pm\bar{3}m$ structures. The metal–insulator transition in ionic systems appears to miss a structural fingerprint.

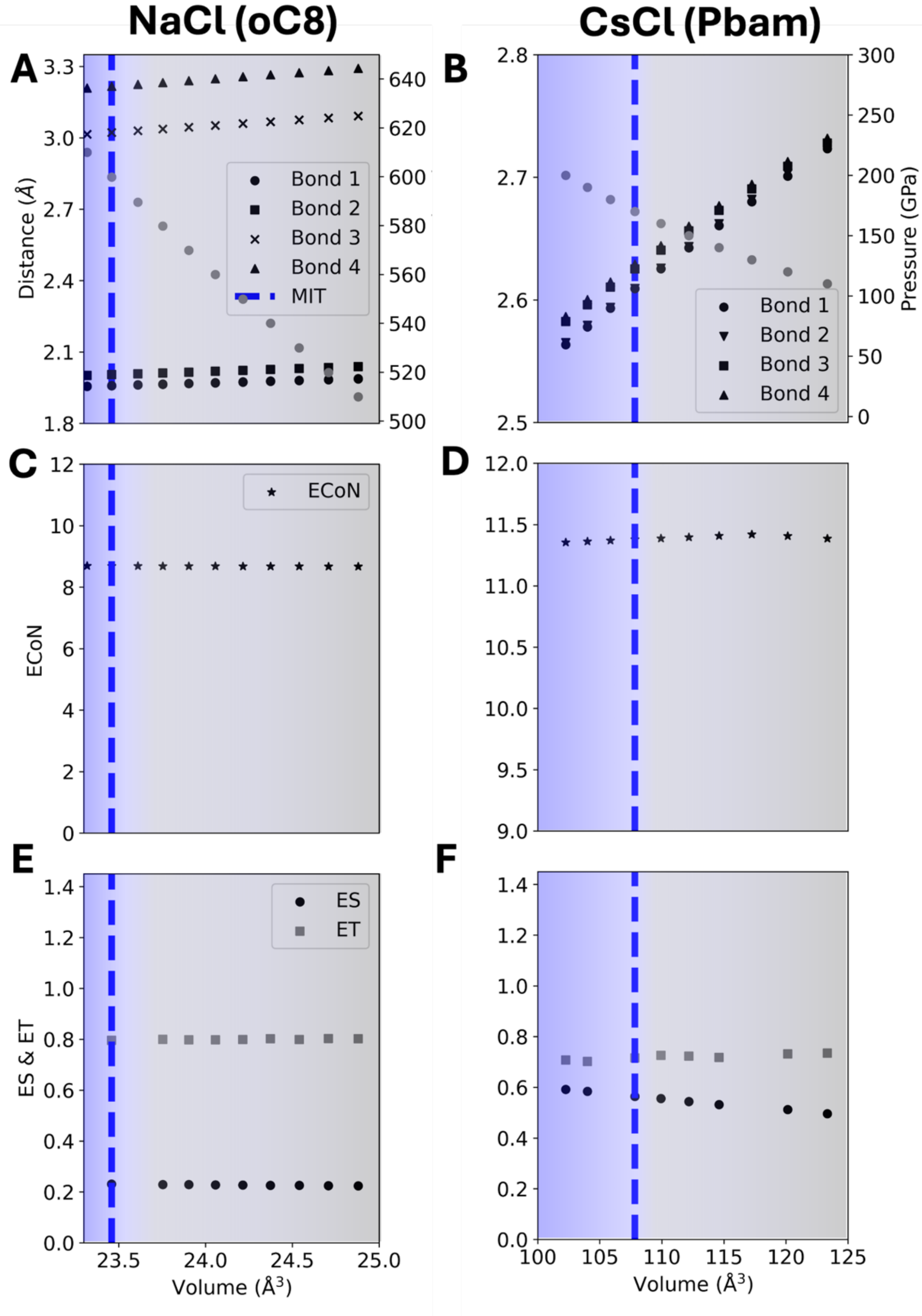

***Figure S8:*** *The Figure presents the evolution of the crystal structure (A and B), ECoN (C and D), and bonding characteristics (E and F) for the predicted high-pressure phases of NaCl and CsCl. The vertical blue line indicates the metal–insulator transition (MIT). The pressures at which metallization is reported to occur are 584 GPa for NaCl[13] and 170 GPa for CsCl[43]. Both values are in good agreement with the calculated MIT pressures.*

## *2.2 Structure of Covalent Solids*

The structural transitions of Ge, Si, and GaAs are discussed first. In the zincblende/diamond-type phase, the bond lengths within the tetrahedral environment of the primitive unit cell were evaluated (Figure S9 A–C). In the insulating phase, shown on the right-hand side of the vertical blue line, the bond lengths decrease approximately linearly upon compression, with only comparatively small changes observed. All compounds exhibit an effective coordination number of 4 (Figure S9 D–F), in agreement with the 8−N rule for semiconductors. The number of electrons shared between neighboring atoms is approximately 1.7 (Figure S9 G–I) and depicts a small increase with compression. The first-order phase transition to the metallic phase occurs at 14.55 GPa for GaAs and at 9.37 GPa and 8.34 GPa for Si and Ge, respectively. The metallic phases of the III–V and group-IV compounds exhibit greater structural diversity than their corresponding ground-state structures. GaAs undergoes a transition to the rocksalt structure ($Fm\bar{3}m$), whereas Si and Ge transform into the β-Sn structure (I41/amd). These first-order phase transitions are accompanied by discontinuous changes in volume, as reflected by the discontinuities along the x-axis in Figure S9.

Metallic GaAs is stable in the $Fm\bar{3}m$ phase, where each constituent atom, for example Ga, is octahedrally coordinated by atoms of the other species. Although the volume decreases by approximately 20%, the bond lengths become larger due to the change from tetrahedral to octahedral coordination. This results in six equivalent bond distances, which decrease approximately linearly with increasing pressure. Importantly, immediately before the transition to the metallic phase, for example at the last point on the right-hand side of the vertical blue line, no indication of a structural change can be observed. The number of shared electrons decreases to approximately 0.9 in the $Fm\bar{3}m$ phase, whereas the number of transferred electrons shows only a minor decrease.

The structural transition of Si and Ge to the β-Sn phase results in different bond lengths in the metallic structure. Four short bonds remain present and slightly increase in length, while two additional longer bonds appear, as indicated by the black triangles in Figure S9 B & C. Further bonding contributions are also present, but are not shown in Figure S9, as they do not provide additional relevant information. Overall, the coordination number increases, leading to a reduction in the number of electrons shared per individual bond.

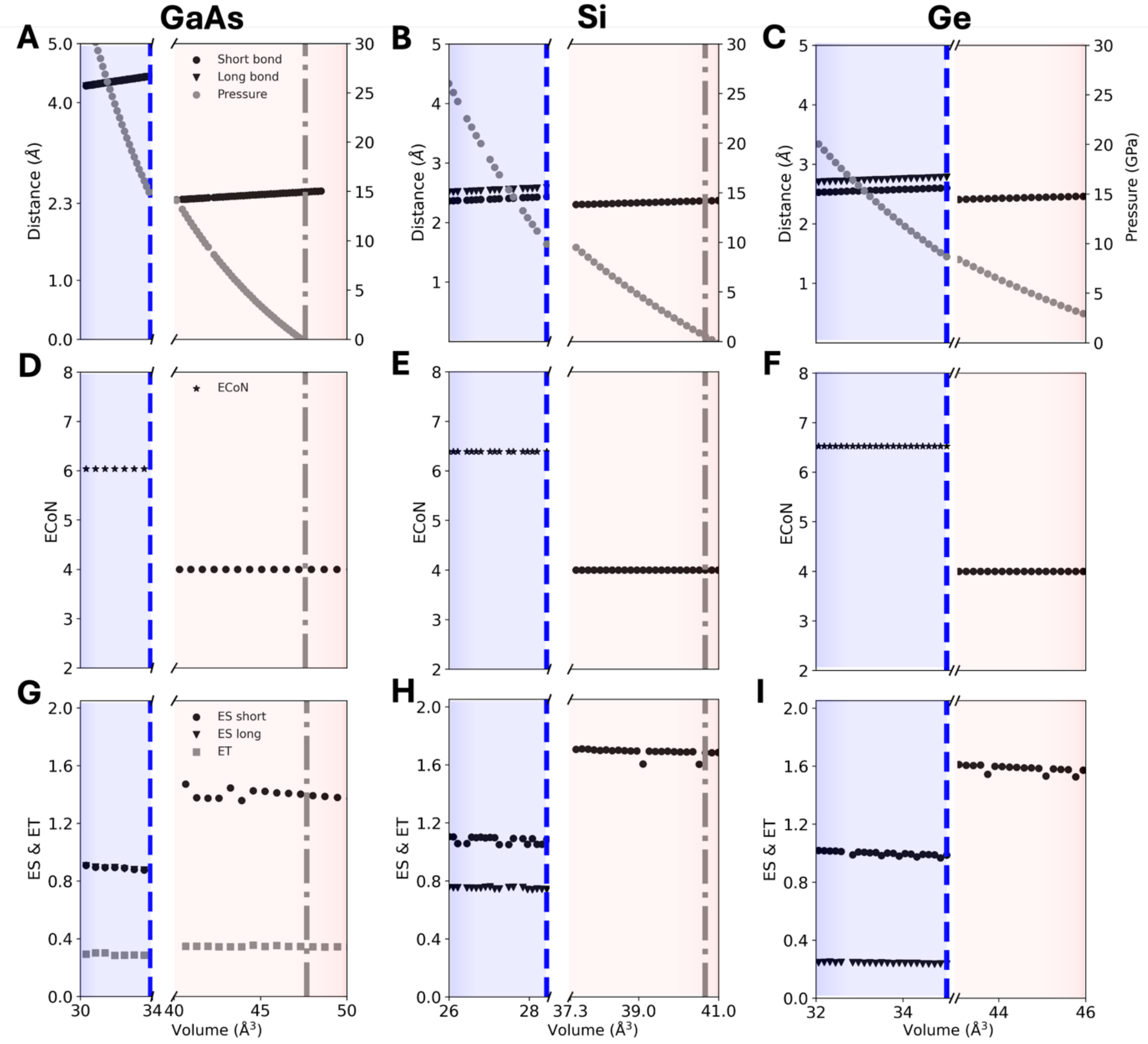

1 **Figure S9:** *Evolution of structure, ECoN, chemical bonding for GaAs, Si and Ge. To characterize the structural progression the nearest neighbor bond lengths as well as the effective coordination number (ECoN) is plotted. Additionally, the number of electrons shared and transferred is depicted for the corresponding bonds.*

## 2.3 Structure of Metavalent Solids:

The structural evolution of metavalent compounds under pressure also exhibits a characteristic behavior related to the suppression of the Peierls distortion. In all systems, three short and three long bonds are present in the distorted phase (Figure S10 A–C). Upon compression, the short and long bond lengths gradually converge until the transition from R3m to Fm3m is completed. In GeTe, SnTe, and GeSe, this transition is accompanied by an MIT, as indicated by the vertical blue dashed line in Figure S10 A–C. At this point, as in GeTe, the ECoN changes from non-integer values to a value of 6 (Figure S10 D–F).

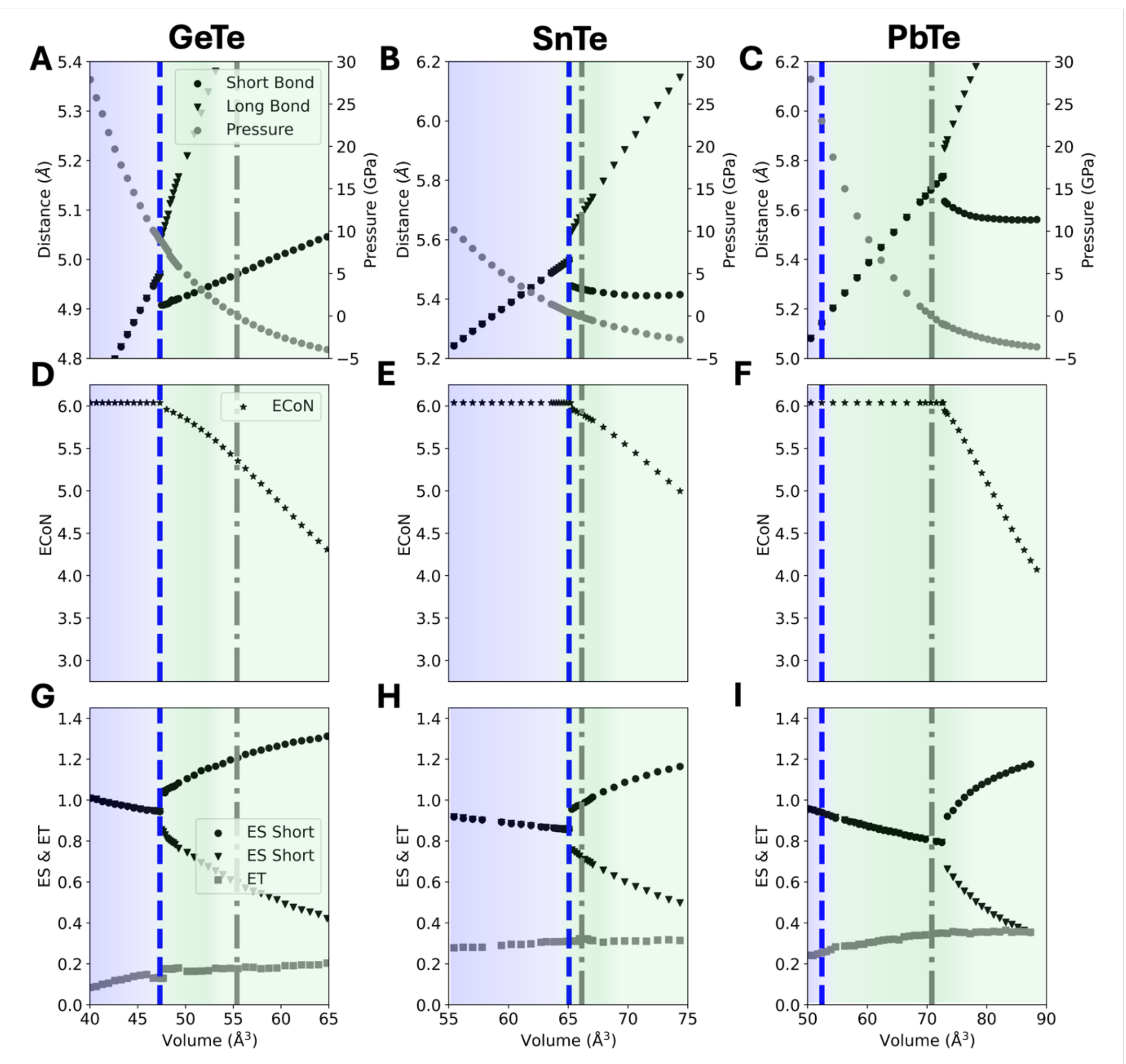

***Figure S10:*** *Evolution of the three short and three long bonds during compression in the R3m and Fm3m phases of GeTe, SnTe, and PbTe (A–C). The corresponding pressures are shown on the right-hand side of each graph and are indicated by grey dots. Panels D–F show the evolution of the ECoN during suppression of the Peierls distortion and in the cubic phase, revealing a progression from non-integer values toward an ECoN of 6. Furthermore, the numbers of electrons shared in the corresponding short and long bonds are evaluated and plotted in panels G–I, together with the number of transferred electrons, indicated by grey squares. The vertical blue line marks the position of the MIT, while the vertical grey line indicates the equilibrium structure of each material.*

# 3 Influence of Ionicity:

## 3.1 GaN, AlN and InN:

As indicated in Figure S11 GaN, AlN, and InN do not become metallic after the first-order phase transition to their respective high-pressure $Fm\bar{3}m$ structures. For these compounds, only a few studies have specifically addressed the subsequent pressure-induced MIT within the $Fm\bar{3}m$ phase. GaN, for example, has been reported to undergo continuous metallization through band-gap closure at approximately 180 GPa[48]. This would classify the MIT of GaN, and possibly also those of AlN and InN, as ionic MITs. This behavior represents a boundary case. In the prototypical covalent and metavalent transitions discussed above, the structural transition "S", defined as the attainment of the high-symmetry, high coordination structure, coincides with the electronic transition "E", defined by band gap closure. Thus, for these cases S = E. The route to E, however, differs and is reached continuously in metavalent compounds, whereas in covalent semiconductors it is reached discontinuously through a first-order transition.

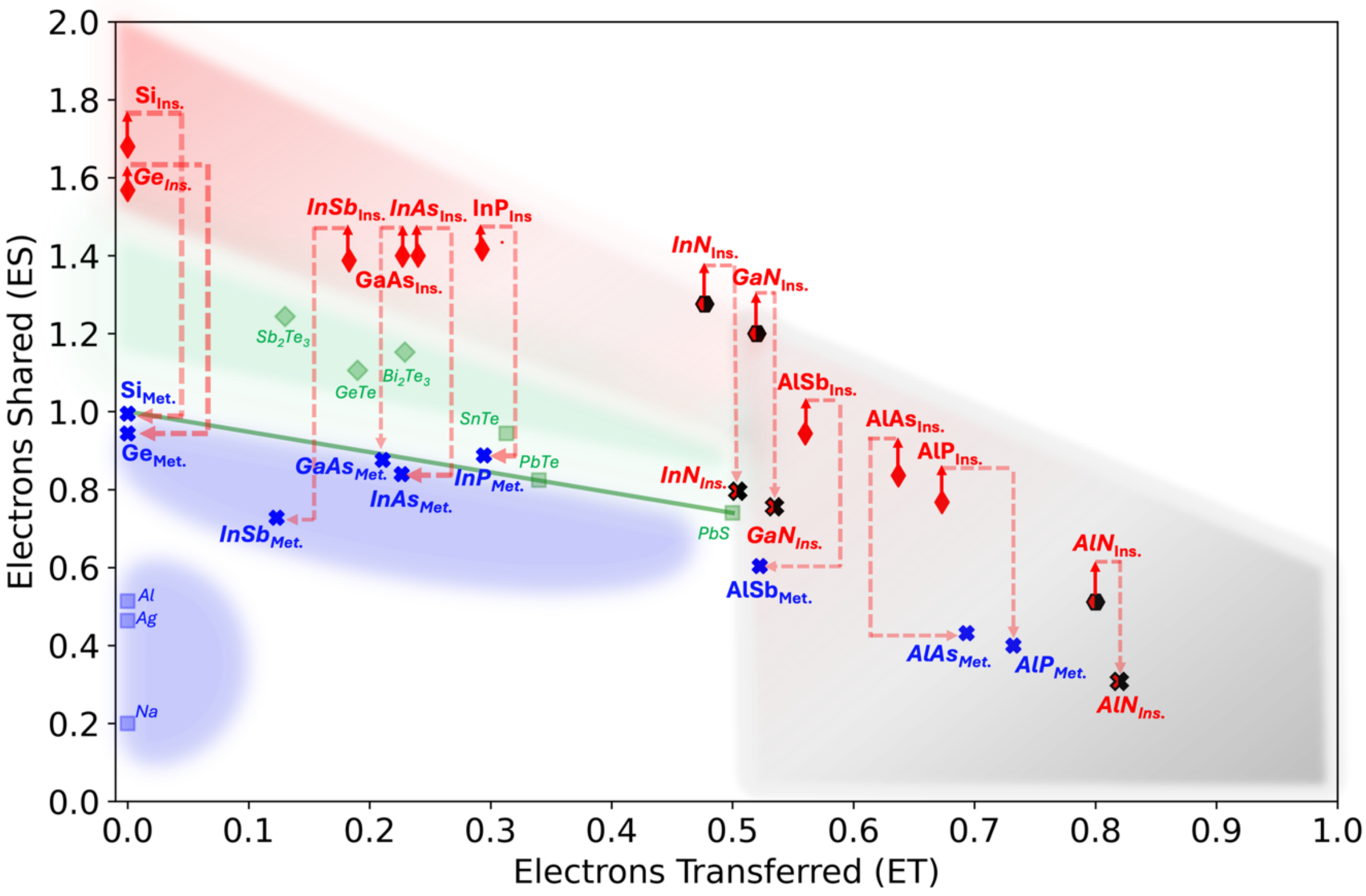

***Figure S11:*** *2D map of chemical bonding in solids depicting pressure-induced MITs for III–V compounds. The map is spanned by two quantum-chemical bonding descriptors: the number of electrons shared between adjacent atoms and the number of electrons transferred, normalized by the oxidation state. The equilibrium structures of the covalent compounds stabilized in the zincblende phase are indicated by red ◊ symbols with the subscript "Ins." InN, GaN, and AlN crystallize in the wurtzite structure, which is indicated by colored ⬡ symbols. Under compression, all compounds show a slight increase in their ES values, indicated by upward-*

*pointing arrows (↑). After a pressure-induced phase transition, the new position on the map is indicated by a blue ╳ symbol with the subscript "Met" if the compound shows no band gap after the transition.*

As discussed in the main manuscript increasing ionicity can decouple these two events, such that the high-coordination structure is reached while a finite band gap remains. This behavior is described by the Phillips-Van Vechten[49] description of the band gap

$$E_g^2 = E_h^2 + C^2$$

Because only limited information is available on the stable structures beyond the wurtzite-to-rocksalt transition[50], a more detailed analysis of these compounds is omitted at this stage. They are therefore classified as examples of polar-covalent compounds with an ionic-like MIT.

## *3.2 PbTe:*

PbTe represents a special case of metavalent bonding because, despite adopting a cubic structure, it retains its insulating character due to its relatively high ionicity[51]. Based on the classification schemes established in recent publications[23,27], PbTe can be classified as metavalently bonded both at its equilibrium structure and throughout its evolution toward the metallic state. This classification is supported by its large Born effective charge, high dielectric constant, and large mode-specific Grüneisen parameter.

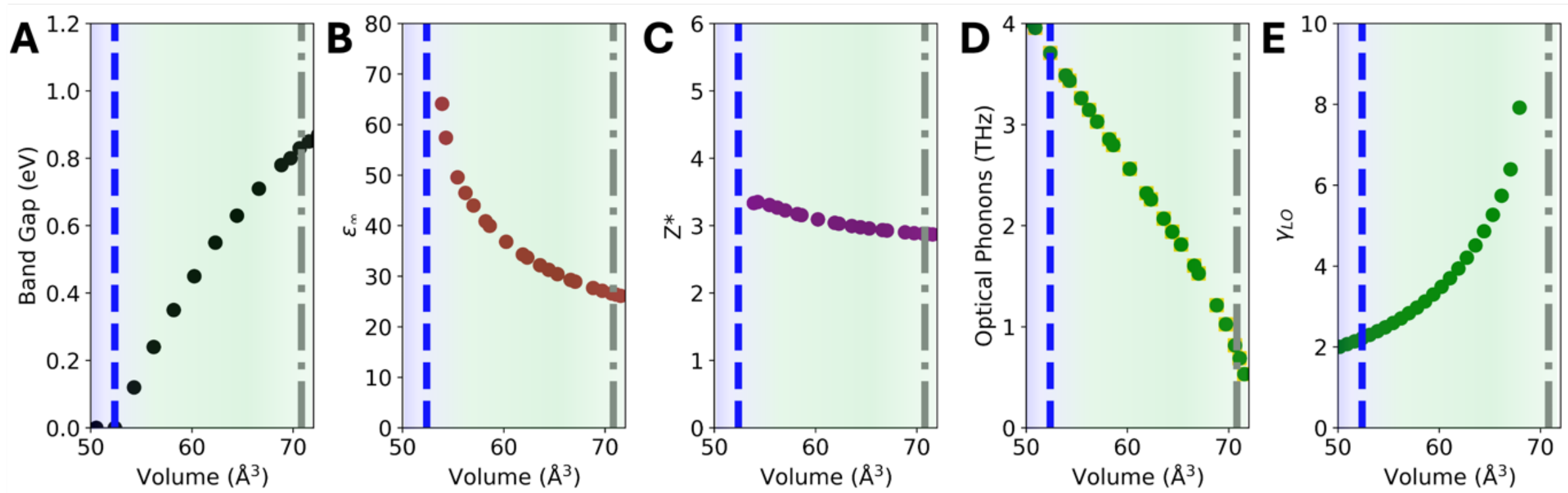

**Figure S12:** Progression of properties between equilibrium structure and metallic state of PbTe. The properties can be categorized as characteristic properties for metavalent compounds, depicting an elevated value of $\varepsilon_\infty$ (B), high Born effective charge (C), soft phonon mode (D), High anharmonicity as described by the mode specific Grüneisen parameter (E). The evolution of these properties at the MIT however progress like systems like NaCl, CsCl and CsF with an increase of $\varepsilon_\infty$ , an unaffected Z* , continuous $\omega_{LO}$ & $\gamma_{LO}$

Focusing on the range between the equilibrium structure and the metallic state (Figure S12 A–E), the classification scheme developed in the main manuscript indicates that PbTe undergoes an ionic metal–insulator transition. The band gap closes continuously, while $\varepsilon_\infty$ increases. At the same time $Z^*_+$ remains nearly unchanged, and both $\omega_{LO}$ and $\gamma_{LO}$ evolve continuously.

Similar behavior is observed in covalent compounds with an increased ionic contribution, such as GaN, AlN, and InN. In these systems, sufficiently high ionicity appears to preserve a finite band gap after the first pressure-induced structural transition, such that further compression is required to close the gap completely[48].

# 4 Progression towards the MIT using QTAIM:

## 4.1 Ionic Compounds:

The progression toward metallization can also be visualized using the two-dimensional bonding map as motivated in the main manuscript. Ionic solids exhibit a unique approach toward the metallic state, which is depicted in Figure S13. Here, for both compounds, all phase transitions toward the metallic high-pressure structures are shown. NaCl undergoes a transition from $Fm3m_{Ins.}$ → $Pm3m_{Ins.}$ → $oC8_{Ins.}$ → $oC8_{Met}$ , CsCl undergoes a $Pm3m_{Ins.}$ → $Pmna_{Ins.}$ → $Pbam_{Ins.}$ → $Pbam_{Met}$ transition series. All transitions except the metal–insulator transition are first-order phase transitions, as indicated by the dotted (for CsCl) or dashed lines (for NaCl) in Figure S13 A. In contrast to covalent and metavalent solids, ionic materials continuously increase their amount of electron sharing during both the first-order phase transitions and the metal–insulator transition. Especially in the phase in which the compounds metallize, the number of shared electrons increases continuously, as shown in Figure S13 B for CsCl and Figure S13 C for NaCl.

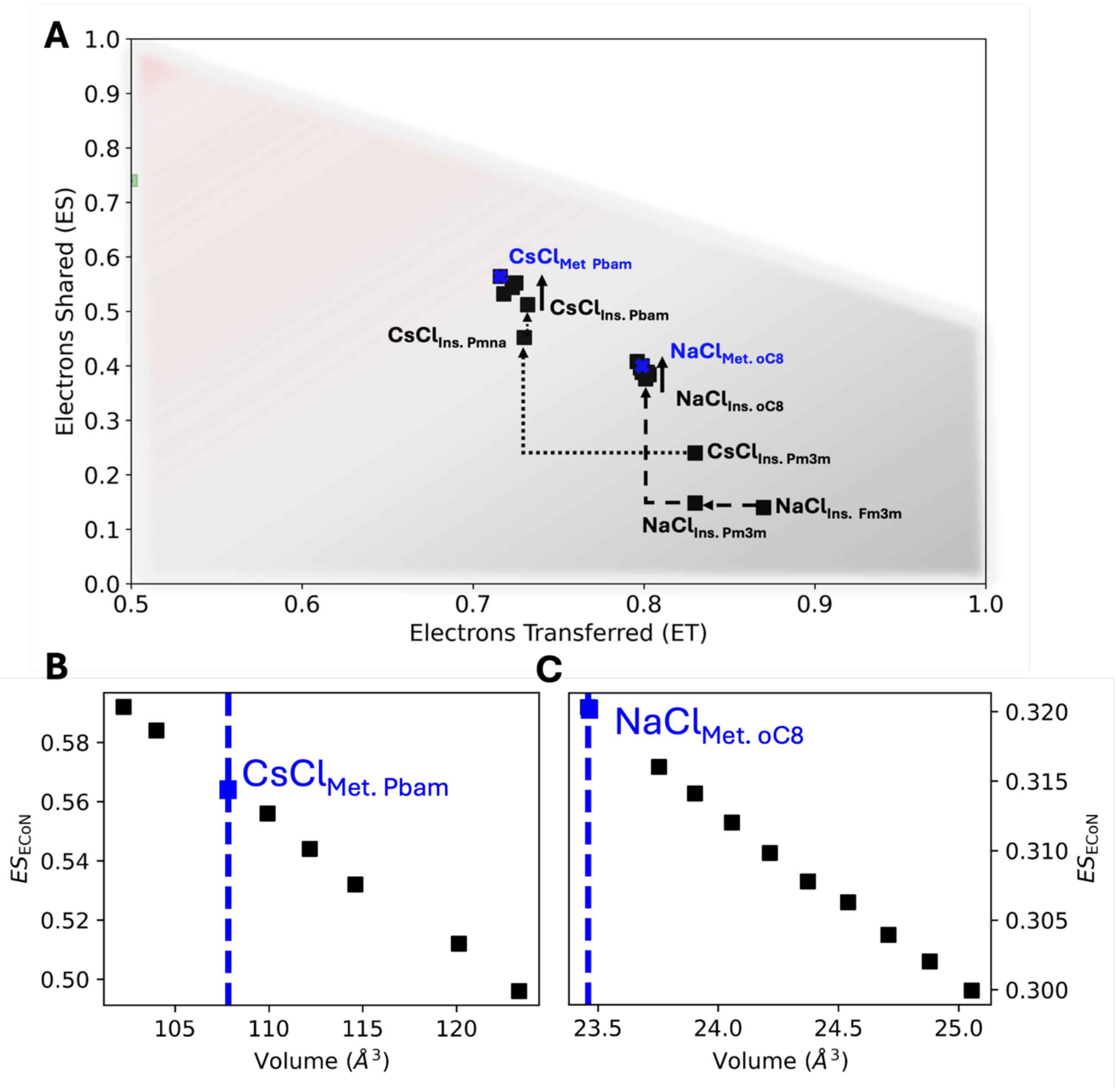

***Figure S13:*** *(A) depicts the values of ECoN weighted ES for all ionic phases including insulating phases based on[13]. The metallic and insulating phases are denoted by the subscript "Met." Or "Ins". First-order transitions between different insulating phases are indicated by a dashed line (NaCl) and a dotted line (CsCl). The transition to the metallic state is represented by a solid arrow, indicating the continuous metal–insulator transition. Each intermediate insulating phase is only represented by a single data point. Among the materials investigated in this study, ionic materials are unique in exhibiting a continuous increase in electron sharing during compression. (B) and (C) depict the ECoN-weighted electron sharing (ES) for CsCl and NaCl in the Pbam and oC8 phases, respectively, showing a continuous increase.*

## *4.2 Covalent Compounds:*

To compare changes in electron sharing between the different structures, ECoN-weighted ES values were calculated and are compared in the following. Assuming that the total number of shared electrons is conserved, $ES_{ins.} \times ECoN_{ins.}=$ $ES_{met.} \times ECoN_{met}$ the number of shared electrons was compared with the corresponding change in ECoN, as summarized in Table S3.

***Table S3:*** *The table presents the number of shared electrons and the effective coordination number in the equilibrium structure, shown in columns 2 and 3, and in the metallic high-pressure structure, shown in columns 4 and 5. These values characterize the changes in chemical bonding and structure across the phase transition.*

| Compound | $ES_{Ins.}$ | $ECoN_{Ins.}$ | $ES_{Met.}$ | $ECoN_{Met.}$ | $Es_{Ins.}/Es_{Met.}$ | $ECoN_{Met.}/ECoN_{Ins.}$ |
|---|---|---|---|---|---|---|
| **Si** | 1.68 | 4 | 1.00 | 6.39 | 1.69 | 1.60 |
| **Ge** | 1.57 | 4 | 0.94 | 6.53 | 1.67 | 1.63 |
| | | | | | | |
| **InSb** | 1.39 | 4 | 0.73 | 7.43 | 1.90 | 1.86 |
| **InAs** | 1.40 | 4 | 0.84 | 6.04 | 1.67 | 1.51 |
| **InP** | 1.42 | 4 | 0.89 | 6.04 | 1.60 | 1.51 |
| **InN** | 1.28 | | | 6.00 | | |
| | | | | | | |
| **AlSb** | 0.94 | 4 | 0.60 | 6.51 | 1.57 | 1.63 |
| **AlAs** | 0.84 | 4 | 0.43 | 6.64 | 1.95 | 1.66 |
| **AlP** | 0.77 | 4 | 0.40 | 6.62 | 1.93 | 1.65 |
| **AlN** | 0.51 | | | | | |
| | | | | | | |
| **GaAs** | 1.40 | 4 | 0.88 | 6.00 | 1.59 | 1.50 |
| **GaN** | 1.20 | | | | | |

Using the two-dimensional bonding map spanned by the number of shared and transferred electrons as a descriptor for chemical bonding, it becomes apparent that the metallic high-pressure structures of the III–V and group-IV semiconductors are located on the metallic side of the demarcation line separating metavalent from metallic bonding (compare Figure S11). The transition from the covalent to the metallic region is always discontinuous. Initially, the number of shared electrons increases, as indicated by the solid arrow above the red region. The first-order insulator-to-metal transition then occurs and is accompanied by discontinuous changes in ES and ECoN. This results in a jump in the bonding map, indicated by the dashed lines, into the metallic region as visible in Figure S11.

# *Data for Figure 6*

Figure 6 in the main manuscript is a combination of data obtained from DFT calculations, literature and experimental measurements conducted at the I. Physikalisches Institut (IA) of the RWTH-Aachen. Table S4 depicts the data used to plot Figure 6, in addition in contains the references used for the conductivity values used.

**Table S4:** The table presents all data points shown in Figure 6, together with the corresponding references[52-58] for the conductivity values. Entries marked with "D" refer to in-house measurements.

| Materials | Conductivity (S/cm) | Ref. | $E_s/E_t$ | Z* | $F_s$ | PME |
|---|---|---|---|---|---|---|
| GeTe | 5.00E+03 | D | 1.24/0.38 | 5.98 | 1.01 | 68 |
| PbTe | 9.80E+03 | 59 | 0.80/0.68 | 5.76 | – | 67 |
| SnTe | 2.90E+03 | 59 | 0.94/0.63 | 6.65 | – | 71 |
| Sb | 2.50E+04 | D | 1.55/0.00 | – | 0.74 | 80 |
| PbS | 1.18E+02 | D | 0.74/1.00 | 4.37 | – | 60 |
| PbSe | 2.40E+02 | D | 0.76/0.86 | 4.80 | – | 65 |
| $Sb_2Te_3$ | 2.30E+03 | D | 1.13/0.40 | 5.93 | – | 82 |
| $Bi_2Te_3$ | 6.60E+02 | 60 | 1.18/0.69 | 6.91 | 0.42 | 76 |
| GaAs | 1.00E−08 | 61 | 1.40/0.68 | 2.20 | – | 3 |
| Si | 1.50E−08 | 62 | 1.68/0.00 | – | – | 9 |
| Ge | 3.30E−02 | D | 1.57/0.00 | – | – | 5 |
| InSb | 2.20E+02 | 63 | 1.39/0.55 | 2.50 | 0.02 | 9 |
| GaSb | 2.70E+02 | 61 | 1.44/0.30 | 2.91 | – | 12 |
| SnSe | 2.50E−05 | D | 1.24/0.83 | 3.49 | 0.03 | 27 |
| InAs | 5.00E+01 | 61 | 1.40/0.72 | 2.74 | 0.05 | – |
| Al | 3.5E+05 | 64 | 0.51/0.00 | – | – | 8 |
| Ag | 6.8E+05 | 65 | 0.49/0.00 | – | – | 2 |
| Zn | 1.70E+05 | D | 0.61/0.00 | – | 0.02 | – |

# 5 *GeSe:*

GeSe is known to crystalize in the Pnma phase under ambient conditions[59]. As a p-bonded compound, GeSe represents an important class of covalent materials distinct from the $sp^3$ bonded III-V compounds. It is therefore an interesting candidate for classification within the framework of this study, particularly with respect to its MIT.

Under pressure, GeSe undergoes a sequence of structural transitions from Orthorhombic (Pnma) → Rhombohedral (R3m) → Cubic ($Fm\overline{3}m$)[60]. The transition from Orthorhombic → Rhombohedral has been reported to be accompanied by a change in chemical bonding from covalent → metavalent [61,62]. In the following paragraphs, the transition toward the metallic state in set into perspective to the Covalent → metavalent transition.

To trace the phase transitions the ground state energies of different GeSe phases were calculated as a function of decreasing volume. Figure S14 presents the results of these calculations and depicts the evolution of enthalpy along the transition sequence from Pnma → R3m → Fm3m. Both in the present calculations and in the cited literature[60], DFT incorrectly predicted the R3m phase as the ground state instead of the experimentally observed Pnma phase[59]. Within the framework of this study, the transition pressure for the Pnma → R3m transition is predicted to be -1.15 GPa, whereas the R3m to Fm3m transition is predicted to occur at approximately at 6.38 GPa. These values are in alignment with previously reported DFT results, which predict transition pressures of approximately $P_{Pnma\rightarrow R3m} \approx -1$ GPa and $P_{R3m\rightarrow Fm3m} \approx 6$ GPa. One possible reason for the discrepancy between the theoretically predicted and experimentally observed ground states is that GeSe and related compounds exhibit a shallow energy landscape, resulting in only small energy differences between competing structural phases.

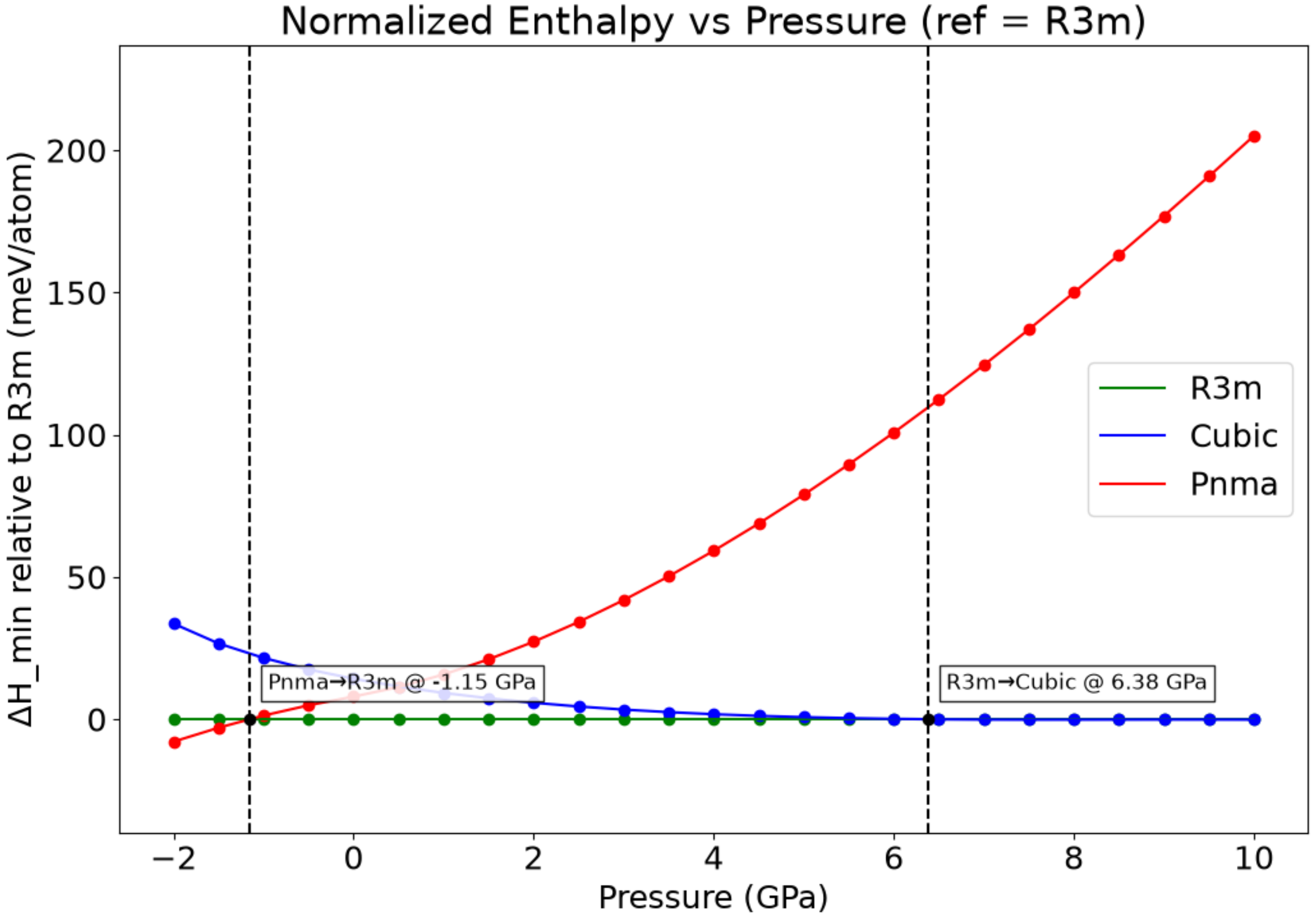

***Figure S14:*** *Enthalpy differences of the competing crystal structures. Pnma phase is indicated by a red line, R3m colored green and the Fm3m phase is colored blue. The transition from Pnma → R3m occurs by a external pressure of -1.15 GPa. The transition to the cubic Fm3m structure occurs at a pressure of 6.38 GPa. The transition pressures are in alignment with trends from literature.*

To categorize the transition based on the evolution of the relevant physical properties, the progression of $E_g$, $\varepsilon_\infty$, $Z_+^*$, $\omega_{LO}$, $\gamma_{LO}$, was calculated and is depicted in Figure S15 A-E. The Pnma phase is indicated by a red background. Comparing these properties with the classification scheme developed elsewhere[27], the Pnma phase can be categorized as covalent. The Pnma → R3m is discontinuous in volume and has already been described by a covalent → metavalent transition. Here, this change in bonding character is reflected by discontinuous changes in the calculated physical properties across the structural transition.

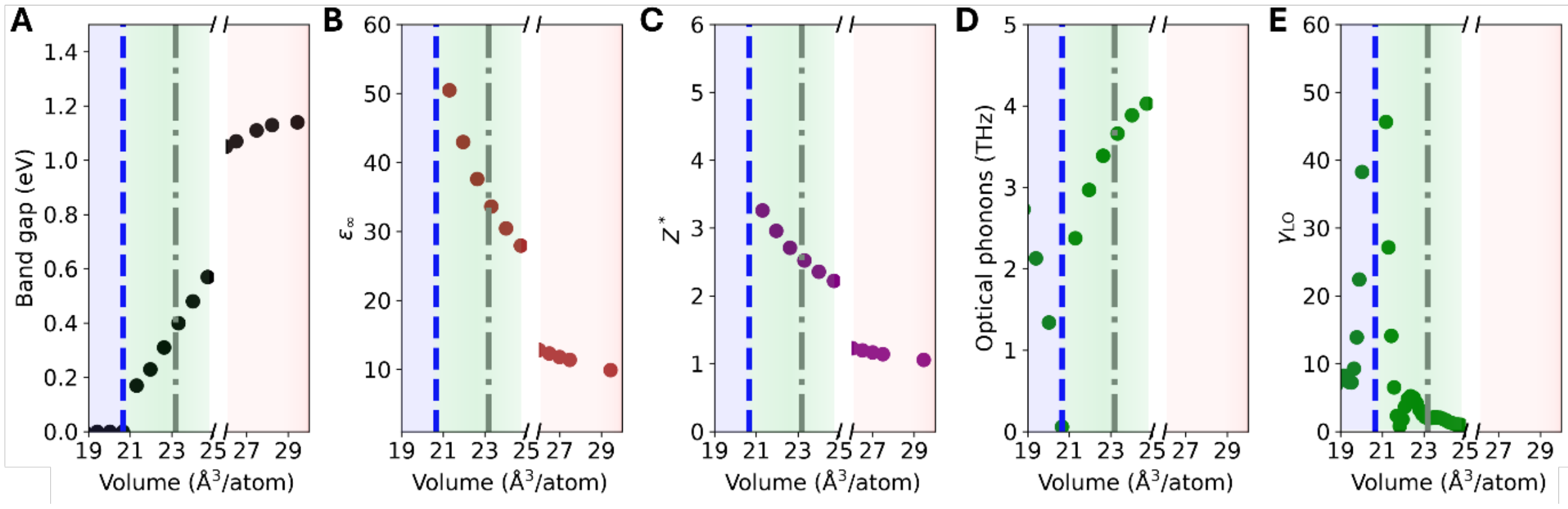

**Figure S15:** Evolution of properties for GeSe in the Pnma (red), R3m (Green) and Fm3m (Blue) phases. (A) depicts the evolution of the bandgap for all three phases. (B) Shows the evolution of $\varepsilon_\infty$. (C) shows the evolution of Z*. The phonon frequencies (d) and the corresponding Grüneisen parameter (e) are only shown for the R3m and Fm3m phase.

Continuing along the transition from R3m → Fm3m, the system exhibits a property evolution comparable to that observed for other metavalent compounds. This behavior is associated with a continuous reduction of the Peierls distortion until it vanishes completely, at which point the system becomes metallic in the Fm3m phase. During this transition $\omega_{LO} \rightarrow 0$ and $\gamma_{LO}$ diverges. The transition from R3m to Fm3m appears to be continuous.

Using the number of electrons shared and electrons transferred, the different phase transitions can be represented on the quantum materials map. In the Pnma phase, ES increases upon compression, as indicated by the solid red arrow in Figure S16. The transition from Pnma → R3m is characterized by a discontinuous jump into the metavalent region of the map, indicated by the dashed red arrow in Figure S16. Within the R3m phase, the number of electrons shared begins to decrease with increasing compression, and the material moves toward the metallic regime, as indicated by the green arrow. The transition from R3m → Fm3m is continuous and is accompanied by a MIT, marked by a blue cross close to the MVB line in Figure S16.

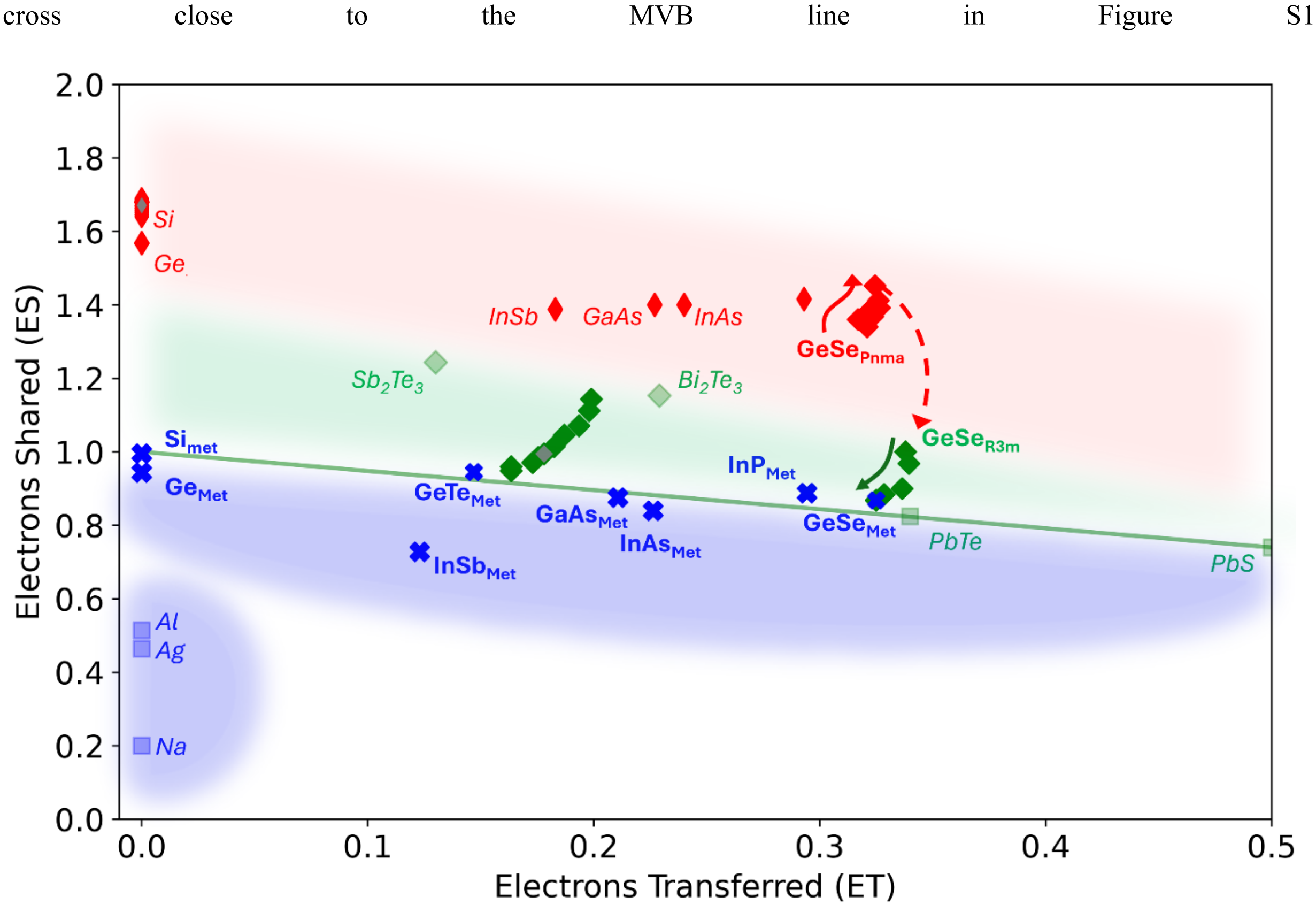

***Figure S16:*** *Quantum-chemical map based on the number of shared electrons, normalized by the effective coordination number (ECoN) of each compound, and the number of transferred electrons. The insulating ground-state and metallic phases of the materials investigated in this work are shown. In particular, the Pnma → R3m → Fm3̄m phase transition of GeSe is highlighted and is discussed in the main text.*

# 6 Summary of Classification:

The results of the discussion of different pathways toward the metallic state in compounds with different types of chemical bonding are summarized in the table S5. The color coding of the compounds in the table is based on a classification scheme developed elsewhere[27] and reflects the bonding mechanism in the equilibrium (zero-pressure) structure. The compounds are sorted according to their pressure-induced metal–insulator transition in the respective columns.

The “covalent MIT” column contains III–V and II–VI compounds, which undergo a first-order phase transition to the metallic state. GeTe and SnTe are compounds that undergo a “metavalent MIT”, which is continuous and exhibits a unique evolution of electronic and vibronic properties. Besides GeTe and SnTe, GeSe also appears in this column. In its ground state, GeSe is stabilized in a covalent orthorhombic Pnma structure, which undergoes a discontinuous transition to the rhombohedral R3m phase that has been shown to exhibit metavalent bonding[61,62]. From this R3m phase, GeSe continuously approaches the metallic state. It is likely that related compounds such as GeS, SnSe, and SnS undergo a similar transition to the metallic state.

The "ionic MIT" column is populated mainly by alkali halides. For these compounds, it has been shown that the MIT is not associated with structural changes, which is attributed to the ionic contributions to the band gap. Interestingly, an ionic-type MIT is also found in some metavalent and covalent compounds that exhibit larger ionic contributions.

The analysis of these compounds revealed how different contributions to the band gap, arising from distinct bonding mechanisms, vanish under pressure. Covalent structures were found to undergo a discontinuous band-gap closure, without a continuous structural evolution. In contrast, metavalent compounds exhibit a continuous transition in which the electronic structure, vibronic properties, and crystal structure are strongly coupled. Compounds with significant ionic contributions to the band gap, which persist even after reaching a high-symmetry phase, display a characteristic transition marked by the absence of structural contributions and a continuous closing of the band gap.

***Table S5:*** *Materials investigated in the framework of this work, in addition to materials sourced from the literature. The color of the compounds represents the bonding type in the ground state.*

| | **Ionic MIT** | **Metavalent MIT** | **Covalent MIT** |
|---|---|---|---|
| **Solid** | NaCl (Pm3m), CsCl (Pm3m), CsF (Pm3m), NaCl (oC8)[13], CsCl (Pbam)[43], CsI[44], CsBr[43], BaS[45], BaSe[45], BaTe[45], KI[42],RbI[42], GaN[48,50], InN[50], AlN[50], PbTe | **GeTe, SnTe, GeSe** | **GaAs,** Si, Ge, InSb, InAs, InP, AlSb, AlP, AlAs, ZnSe[16,18], ZnTe[16,18], ZnS[16,18] |